%% file: main.tex
\documentclass[acmsmall]{acmart}
\pdfoutput=1
\AtBeginDocument{%
  \providecommand\BibTeX{{%
    \normalfont B\kern-0.5em{\scshape i\kern-0.25em b}\kern-0.8em\TeX}}}

\setcopyright{acmcopyright}
\copyrightyear{2022}
\acmYear{2022}
\acmDOI{XXXXXXX.XXXXXXX}

\usepackage{multirow}
\usepackage{comment}
\usepackage{booktabs}
\usepackage{tablefootnote}
\usepackage{multirow}
\usepackage{amsfonts}
\usepackage{soul}
\usepackage{lipsum}
\usepackage{comment}
\usepackage[linesnumbered,ruled,vlined]{algorithm2e}
\usepackage{bm}
\usepackage{enumitem}
\usepackage{color}
\usepackage{makecell}

\usepackage{pgfplotstable}
\usepgfplotslibrary{statistics}
\usepackage{tikz}
\usepackage{subfigure}
\usepackage{pgfplots}
\pgfplotsset{compat=1.14}
\usetikzlibrary{arrows,shapes,snakes,automata,backgrounds,petri, matrix, patterns}
\usepackage{enumerate}
\usepackage{tikz-qtree}
\usepgfplotslibrary{groupplots,external}
\pgfplotsset{compat=newest}

\usepackage{pifont}

\definecolor{1}{RGB}{141,211,199}
\definecolor{2}{RGB}{255,255,179}
\definecolor{3}{RGB}{190,186,218}
\definecolor{44}{RGB}{251,128,114}
\definecolor{5}{RGB}{156,209,255}
\definecolor{66}{RGB}{253,180,98}
\definecolor{77}{RGB}{179,222,105}
\definecolor{8}{RGB}{251,154,153}
\definecolor{99}{RGB}{31,120,180}

\definecolor{blue}{RGB}{42,85,127}
\definecolor{6}{RGB}{107,94,139}
\definecolor{9}{RGB}{78,129,121}

\definecolor{7}{RGB}{205,135,113}
\definecolor{4}{RGB}{127,66,82}

\begin{document}

\title{A Multi-Channel Next POI Recommendation Framework with Multi-Granularity Check-in Signals}

\author{Zhu Sun}
\email{sunzhuntu@gmail.com}
\affiliation{%
  \institution{Institute of High Performance Computing; Centre for Frontier AI Research, A*STAR}
  \streetaddress{1 Fusionopolis Way}
  \country{Singapore}
  \postcode{138632}
}

\author{Yu Lei}
\authornote{corresponding author}
\email{leiyu0160@gmail.com}
\affiliation{%
  \institution{Yanshan University}
  \streetaddress{438 Hebei Ave}
  \city{Qihuangdao}
  \country{China}
  \postcode{066104}
}

\author{Lu Zhang}
\email{zhang_lu@ntu.edu.sg}
\affiliation{%
  \institution{Nanyang Technological University}
  \streetaddress{50 Nanyang Ave}
  \country{Singapore}
  \postcode{639798}}

\author{Chen Li}
\email{lichen36211@gmail.com}
\affiliation{%
  \institution{Yanshan University}
  \streetaddress{438 Hebei Ave}
  \city{Qinghuangdao}
  \country{China}
  \postcode{066104}
}

\author{Yew-Soon Ong}
\email{ong\_yew\_soon@hq.a-star.edu.sg;    asysong@ntu.edu.sg}
\affiliation{%
 \institution{A*STAR Centre for Frontier AI Research; Nanyang Technological University}
 \streetaddress{1 Fusionopolis Way}
 \country{Singapore}
 \postcode{138632}}

\author{Jie Zhang}
\email{zhangj@ntu.edu.sg}
\affiliation{%
  \institution{Nanyang Technological University}
  \streetaddress{50 Nanyang Ave}
  \country{Singapore}
  \postcode{639798}}




\renewcommand{\shortauthors}{Trovato and Tobin, et al.}

\begin{abstract}
Current study on next POI recommendation mainly explores user sequential transitions with the fine-grained individual-user POI check-in trajectories only, which suffers from the severe check-in data sparsity issue. In fact, coarse-grained signals (i.e., region- and global-level check-ins) in such sparse check-ins would also benefit to augment user preference learning.
Specifically, our data analysis unveils that user movement exhibits noticeable patterns w.r.t. the regions of visited POIs. Meanwhile, the global all-user check-ins can help reflect sequential regularities shared by the crowd. 
We are, therefore, inspired to propose the MCMG: a Multi-Channel next POI recommendation framework with Multi-Granularity signals categorized from two orthogonal
perspectives, i.e., fine-coarse grained check-ins at either POI/region level or local/global level. 
Being equipped with three modules (i.e., global user behavior encoder, local multi-channel encoder, and region-aware weighting strategy), 
MCMG is capable of capturing both fine- and coarse-grained sequential regularities as well as exploring the dynamic impact of multi-channel by differentiating the region check-in patterns. Extensive experiments on four real-world datasets show that our MCMG significantly outperforms state-of-the-art next POI recommendation approaches.
\end{abstract}



\maketitle
\input{section/intro}

\input{section/related}

\input{section/analysis}

\input{section/method}

\input{section/experiments}

\input{section/conclusion}


\section*{Acknowledgment}
This work is supported by A*STAR Center for Frontier Artificial Intelligence Research and in part by the Data Science and Artificial Intelligence Research Centre, School of Computer Science and Engineering at the Nanyang Technological University (NTU), Singapore.

\input{reference.bbl}

\bibliographystyle{ACM-Reference-Format}
\bibliography{reference}


\end{document}

%% file: section/intro.tex
\section{Introduction}
Recently, next point-of-interest (POI) recommendation has been widely studied in both academia and industry~\cite{zhao2020go,huang2019attention,feng2015personalized,lim2020stp,cui2022sequential,zang2022cha}, which aims to recommend next locations (i.e., POIs) to users based on their historical check-in trajectories (i.e., sequences). 
With the assumption that a user's next movement is highly related to his recently visited POIs, current next POI recommendation methods~\cite{zhao2016stellar,kong2018hst, wang2020next,liu2019geo} mainly resort to individual users' POI check-in records for user preference learning. For instance, many sequential-based methods utilize Markov chain~\cite{rendle2010factorizing,cheng2013you}, recurrent neural networks (RNN)~\cite{feng2018deepmove,wu2020personalized} and self-attention~\cite{wang2021spatial,luo2021stan,lin2021pre} to explore the underlying sequential patterns hidden in the individual-user POI check-in 
trajectories, some of which are further facilitated by the spatial~\cite{liu2016predicting,liu2019geo,zhao2020go} or temporal~\cite{zhao2016stellar,liu2016predicting,lim2020stp} contextual information. 

However, our preliminary data analysis shows that the individual users' POI check-in records are extremely sparse, i.e., each user only visits a limited number of POIs during his daily trajectory. For instance, the average number of visited POIs of each trajectory is merely 2 in many cities (e.g., New York, Phoenix, Calgary and Singapore) on Foursquare\footnote{\url{https://foursquare.com/}}. Despite the success of current next POI recommendation methods, we argue that the data sparsity issue heavily restricts their capabilities of mining accurate user behavior patterns based on individual-user POI check-in trajectories only, thus impeding the eventual recommendation performance.

\begin{figure}[t]
\centering
\includegraphics[width=0.5\linewidth]{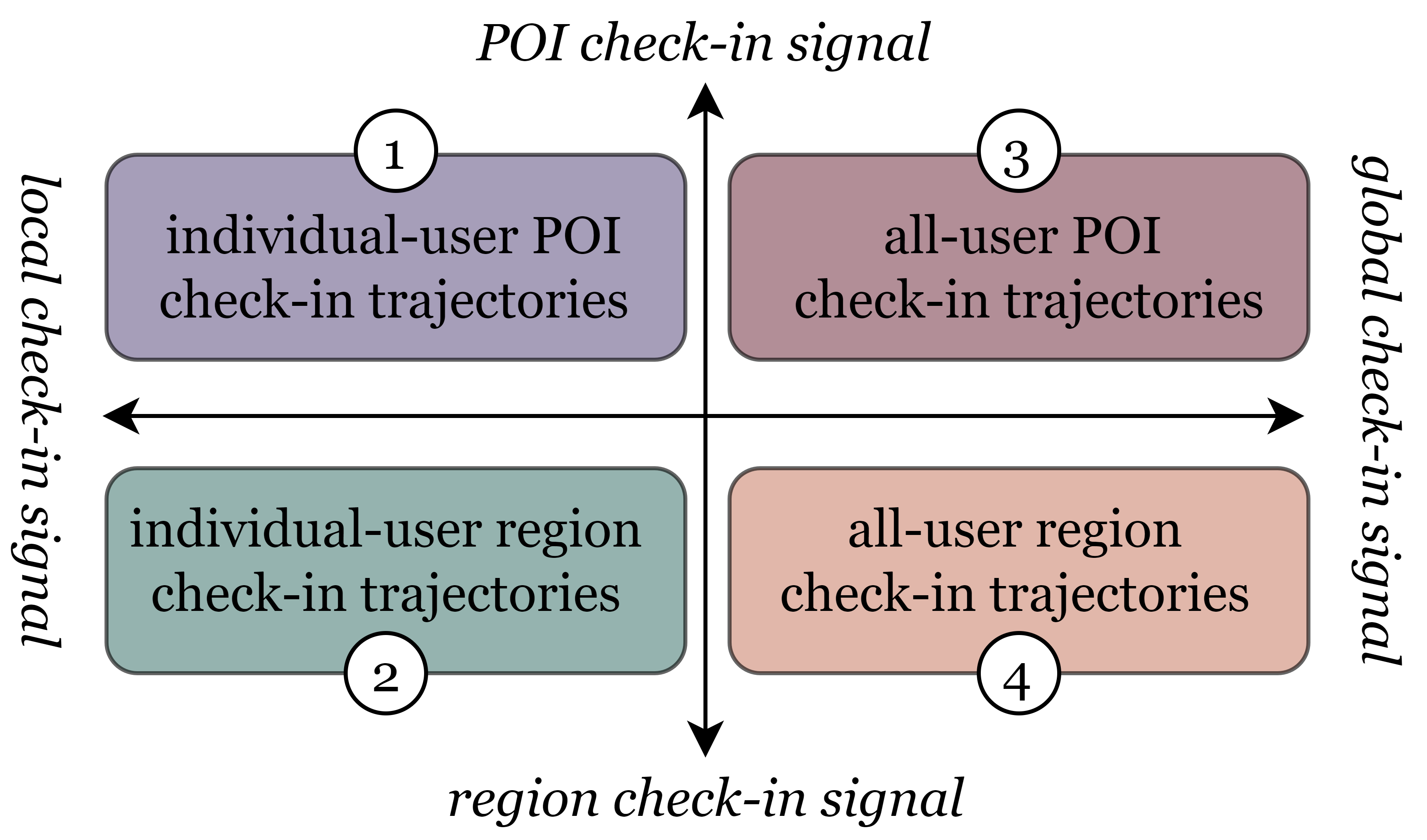}
\caption{The user check-in signals in multi-granularity from two orthogonal perspectives.}\label{fig:information_granularity}
\vspace{-0.1in}
\end{figure}

As a matter of fact, the coarse-grained signals in such sparse check-in records would also help unveil user behavior patterns for augmented user preference learning. In particular, our data analysis in Section~\ref{sec:data_and_analysis} indicates that user movement
exhibits noticeable patterns w.r.t. regions of visited POIs, that is, more than 80\% of users' daily trajectories are within one region across different cities. In this sense, such a signal (i.e., individual-user region check-in trajectories) could greatly benefit for scaling down the search space of candidate POIs to be recommended.
Meanwhile, there are always common sequential regularities among the crowd, for instance, users favor visiting several shops (e.g., HM and Zara) at one time for making the best choice (e.g., shop $\rightleftharpoons$ shop); besides, they would also like to seek food or drink during shopping (i.e., shop $\rightleftharpoons$ food/drink)~\cite{sun2021point}. Consequently, it is much easier to detect common sequential behavior patterns by statistically analyzing all-user POI check-in trajectories for augmented user preference inference.

\begin{figure}[t]
\centering
\includegraphics[width=0.7\linewidth, keepaspectratio]{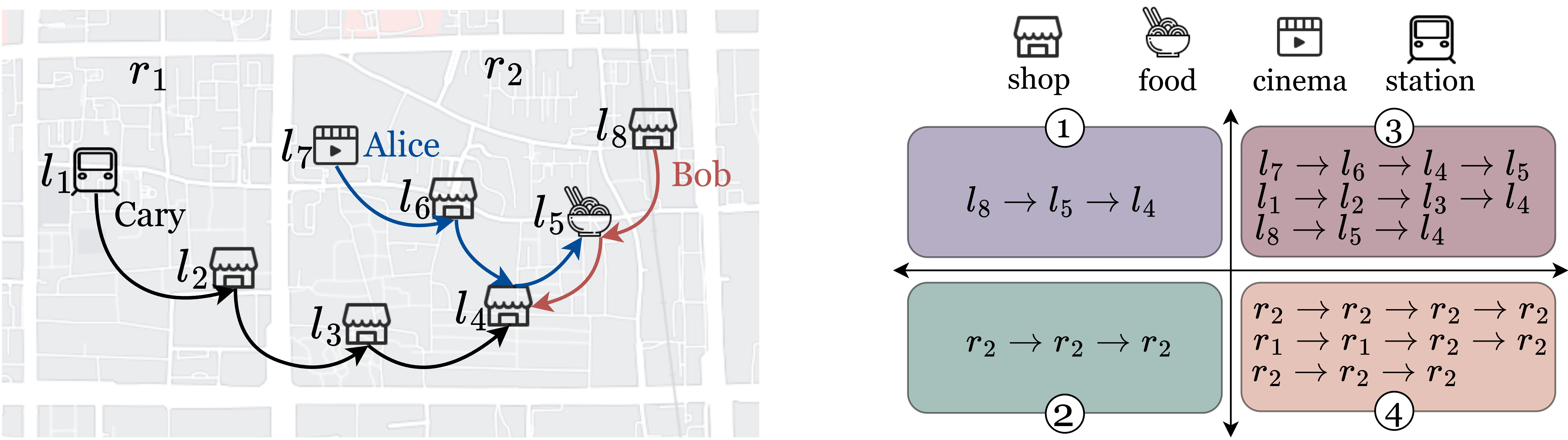}
\caption{A running example of \textcolor{black}{mining} multi-granularity signals from Bob's check-in records. 
The blue, red and black lines are the trajectories of Alice, Bob and Cary, respectively; and the white blocks help divide different regions (i.e., region $r_1$ and region $r_2$).
}\label{fig:running_example}
\vspace{-0.1in}
\end{figure}

As such, we seek to mine multi-granularity signals with two orthogonal angles from users' check-in records. Specifically, as illustrated in Fig.~\ref{fig:information_granularity}, the vertical direction classifies the check-in records into fine-grained POI check-ins (i.e., \textcircled{1} and \textcircled{3}) and coarse-grained region check-ins (i.e., \textcircled{2} and \textcircled{4}); while the horizontal direction is characterized by local individual-user check-ins (i.e., \textcircled{1} and \textcircled{2}) and global all-user check-ins (i.e., \textcircled{3} and \textcircled{4}). 
Accordingly, Fig.~\ref{fig:running_example} depicts a running example of \textcolor{black}{mining} multi-granularity signals from Bob's check-in records. 
When predicting next locations for Bob based on his POI check-in trajectory \textcircled{1}, his region check-in trajectory \textcircled{2} could help navigate the next possible visit in region $r_2$. In addition, based on all-user check-in trajectories \textcircled{3}, we may deduce a common sequential regularity that users are more likely to visit several shops consecutively for diverse choices. Given these clues, we can recommend shops $l_6$ and $l_3$ in region $r_2$ as Bob's next possible check-ins after visiting $l_4$.

There are a few studies attempting to integrate multi-granularity signals to further enhance the accuracy of next POI recommendation. 
They, however, are restricted by merely considering one direction~\cite{li2021discovering,lim2020stp} (i.e., \textcircled{1} and \textcircled{3} in Fig.~\ref{fig:information_granularity}). Besides, other studies considering the vertical direction (i.e., \textcircled{1} and \textcircled{2}) are specially designed only for out-of-town~\cite{yin2016adapting, yin2015joint, yin2017spatial} or general POI recommendation~\cite{zeng2021pr,liu2014exploiting,luo2020spatial} scenarios, rather than next POI recommendation. 
Alternatively stated, how to seamlessly incorporate multi-granularity signals with two orthogonal perspectives from check-in records as in Fig.~\ref{fig:information_granularity} is still remaining unresolved, and worth in-depth exploration for next POI recommendation task. Inspired by this, we thus propose a \underline{M}ulti-\underline{C}hannel next POI recommendation framework with \underline{M}ulti-\underline{G}ranularity check-in signals (MCMG). 

To start, we perform analysis on the global all-user region check-ins (\textcircled{4} in Fig.~\ref{fig:information_granularity}) to gain essential findings 
as design guidance, and then propose MCMG which mainly consists of three modules. 
\begin{itemize}
    \item \textbf{Global User Behavior Encoder}. We construct a directed POI graph via the global all-user POI check-ins (\textcircled{3} in Fig.~\ref{fig:information_granularity}), whereby the graph convolutional network (GCN)~\cite{kipf2016semi} is applied to help capture the augmented sequential regularity 
    \textcolor{black}{from the crowd}, thus learning more expressive POI representations.
    \item \textbf{Local Multi-Channel Encoder}. A three-channel encoder is designed to encode the local check-in signals via the self-attention mechanism~\cite{vaswani2017attention}. Specifically, the main channel -- location encoder aims to capture POI transition patterns in individual-user POI check-ins (\textcircled{1} in Fig.~\ref{fig:information_granularity}), 
    \textcolor{black}{where the input POI representations are learned via the global user behavior encoder;}
    one auxiliary channel -- region encoder facilitates to learn region transition regularity in individual-user region check-ins (\textcircled{2} in Fig.~\ref{fig:information_granularity}); and the other auxiliary channel -- category encoder is inspired by~\cite{zhang2020interactive} to further unveil latent activity transition patterns. 
    \item \textbf{Region-aware Weighting Strategy}. Ultimately, a novel region-aware weighting strategy is proposed to aggregate the three local channels by alleviating potential contradictory predictions. That is, the predicted POI, region and category from the three channels may be mismatched with each other; it is thus essential to dynamically determine the impact of different channels. Intuitively,
    simpler region check-in patterns (e.g., a trajectory with check-ins in a same region only) guarantee more accurate predictions by the region encoder, thus should be more reliable and contribute more than the other encoders on the final next POI prediction, \textit{vice versa}. 
\end{itemize}

In summary, our main contributions lie in three-fold. 
\begin{itemize}
\item We systematically organize and mine the multi-granularity signals with two orthogonal angles from the sparse check-ins, and conduct quantitative analysis to show the usefulness of region context, which provides essential guidance for the next POI recommender design.
\item We are the first to design a multi-channel next POI recommendation framework with multi-granularity check-in signals from two orthogonal perspectives. Most importantly, a novel region-aware weighting strategy is devised to dynamically balance the impact of different channels with different region check-in patterns.
\item We conduct extensive experiments on four real-world datasets to demonstrate that our proposed MCMG significantly outperforms state-of-the-arts, with average lifts of 9.7\% and 14.2\% on HR and NDCG, respectively.
\end{itemize}

%% file: section/related.tex
\section{Related Works}

This section provides an overview of the related state-of-the-art sequential based next POI recommendation and region based POI recommendation.

Recent next POI recommendation approaches adopt RNN and its variants (e.g., LSTM and GRU)~\cite{hochreiter1997long} to help capture the sequential patterns of human movement.
For instance, ST-RNN~\cite{liu2016predicting} extends RNN with time- and distance-specific matrices to model users' spatial and temporal patterns.
SERM~\cite{yao2017serm} jointly learns embeddings of temporal and semantic factors containing information about user preference in a unified RNN framework.
DeepMove~\cite{feng2018deepmove} merges an attention model with RNN to better predict user movements by capturing multi-level periodicity and heterogeneous transition regularity.
ATST-LSTM~\cite{huang2019attention} leverages an attention-based spatial-temporal LSTM to focus on the most relevant check-in records in a trajectory.
Later, STGN~\cite{zhao2020go} equips the LSTM model with two pairs of spatial and temporal gates to capture the long- and short-term interactions that exist in the individual user's POI check-ins. ASPPA~\cite{zhao2020discovering} adopts a state-based stacked RNN to learn users' behavior patterns in a hierarchical manner.
LSTPM~\cite{sun2020go} utilizes a geo-dilated RNN to capture the correlation across non-consecutive POIs and explores the long- and short-term user preferences through historical and present trajectories.


After that, self-attention based methods (e.g., SASRec~\cite{kang2018self} and Bert4Rec~\cite{sun2019bert4rec}) have proliferated in next-item recommendation due to their capability to selectively focus on salient segments for sequential pattern mining.
There are a few self-attention based approaches in next POI recommendation. For instance, 
GeoSAN~\cite{lian2020geography} is a self-attention based model which introduces the hierarchical gridding of each location with a self-attention-based geography encoder to better use the spatial information.
STSAN~\cite{wang2021spatial} and STAN~\cite{luo2021stan} integrate spatial-temporal information with the self-attention for next location prediction.
CTLE~\cite{lin2021pre} is a recent state-of-the-art contextual (i.e., temporal) location embedding method which is built upon a bi-directional Transformer framework.
However, all methods mentioned above merely emphasize the user preference hidden in local individual-user POI check-in trajectories (i.e., \textcircled{1} in Fig.~\ref{fig:information_granularity}),
but ignore the more coarse-grained signals, e.g., global all-user check-ins (\textcircled{3} in Fig.~\ref{fig:information_granularity}) for better user preference inference.


With the great success of graph neural networks (GNNs) in modeling global all-user interactions for item recommendations~\cite{wang2019neural,chen2020revisiting,sun2020neighbor,huang2021position,he2020lightgcn}, 
GNNs has been widely adopted to tackle POI~\cite{chang2020learning} and next POI~\cite{wang2021attentive, lim2020stp,li2021discovering} recommendation problems. 
In particular, 
STP-UDGAT~\cite{lim2020stp} leverages GNN to explore users' global and local behavior patterns with the assistance of geographical, temporal, and frequency information. Later, ASGNN~\cite{wang2021attentive} employs GNN to learn the global information between all POIs and adopts an attention network to acquire users' long- and short-term preferences.
Meanwhile, SGRec~\cite{li2021discovering} equips the graph attention network (GAT) with category information to capture the global pattern between all POIs and explores the local sequential patterns by a left-to-right unidirectional position-aware attention network.
%
%
Although such GNN based next POI recommenders all take into account both local individual-user and global all-user POI check-ins (\textcircled{1} and \textcircled{3} in Fig.~\ref{fig:information_granularity}), they ignore to fully explore multi-granularity signals (e.g., region) in the vertical direction (\textcircled{2} and \textcircled{4} in Fig.~\ref{fig:information_granularity}).

Region, as a coarse-grained geographical attribute of POIs, has been exploited to facilitate different tasks in location based social networks. In particular, \textit{a branch of study utilizes users' trajectory patterns on region for region recommendation}. 
For instance, RegRS~\cite{pham2017general} takes the spatial influence between POIs into consideration for region prediction in new areas. Liu et al.~\cite{liu2018efficient} measure region similarities with the assistance of POI attributes (e.g., geo-location) to improve the performance of region recommendation. 
\textit{Some works employ users' preference on different regions to assist in out-of-town POI recommendation}. Specifically, 
JIM~\cite{yin2015joint} accounts for the effect of temporal, spatial, social, and region popularity of POIs on users' out-of-town spatial mobility.
ST-LDA~\cite{yin2016adapting} employs the social and spatial correlation information to capture users' interest shifts in different regions and learn user preference on out-of-town POIs.
Similarly, SH-CDL~\cite{yin2017spatial} learns hierarchical user preferences via  heterogeneous information (e.g., temporal, spatial and popularity) of POIs, aiming to capture user preference in different regions.
\textit{There are also researchers fusing region information to facilitate general POI recommendation}. Early study~\cite{liu2014exploiting} employs 
two levels (instance- and region-level) of geographical characteristics to estimate user preference on locations.
Later, 
MPR~\cite{luo2020spatial} differentiates the granularity of spatial objects on different levels to infer user preference, where region (i.e. city) is one of the granularity levels.
Recently, 
PR-RCUC~\cite{zeng2021pr} clusters locations into different regions, which are then utilized together with geographical distance and category information to enhance the traditional collaborative filtering model.

However, the region-aware methods mentioned above neglect to unveil the user behavior pattern w.r.t. regions of visited POIs and fail to jointly model both behavior patterns regarding regions and POIs.
To the best of our knowledge, only few works attempt to exploit users' behavior patterns on region for next POI recommendation. Particularly, HST-LSTM~\cite{kong2018hst} mines user movement patterns among various functional zones, so as to predict the next movement, by assuming that users' movements among different functional zones reveal different socioeconomic activities, e.g., shopping, entertaining and working. It is noteworthy that the region mentioned in HST-LSTM is always associated with a specific function, such as education or business. In contrast, the regions in our study refer to the geographically-close areas which may contain POIs with different functions.

Meanwhile, location-aware recommendations usually refer to multiple related prediction tasks, such as region and POI predictions. 
As a result, multi-task learning (MTL)~\cite{caruana1997multitask} has been widely used in location based social networks for various tasks, e.g., travel route planning~\cite{huang2020multi}, transportation~\cite{liu2020incorporating},  and POI recommendation~\cite{xia2020mtpr}. 
Regarding next POI recommendation, MCARNN~\cite{liao2018predicting} utilizes the temporal context to control LSTM to predict users' preferences on location and activity simultaneously. Later, iMTL~\cite{zhang2020interactive} is designed to interactively perform both activity and location prediction tasks with uncertain check-ins. MTNR~\cite{zhong2020multi} jointly learns when and where users are likely to go next. Nevertheless, all the MTL based methods ignore the potential inconsistent prediction results from different tasks, for instance, the predicted location and activity may not always match with each other in MCARNN and iMTL. 
By contrast, we propose a region-aware weighting strategy to help alleviate such contradictory issue by dynamically adapting the impacts of different channels.

%% file: section/analysis.tex
\section{Data Overview and Analysis}\label{sec:data_and_analysis}

\input{plot/region_k-means} 

This section first provides an overview of data utilized in our study and then carries out comprehensive analysis with regards to the coarse-grained region check-in trajectories (\textcircled{4} in Fig.~\ref{fig:information_granularity}). Thereby, several major findings are gained to assist in the model design. 

\input{table/statistics}

\input{plot/analysis_results}

\subsection{Data Overview}\label{subsec:data-overview}
The datasets utilized in our study come from Foursquare~\cite{yang2016participatory} check-in records in four cities, namely Calgary (CAL), Phoenix (PHO), Singapore (SIN) and New York (NY) from April 2012 to September 2013.
Following state-of-the-arts~\cite{zhang2020interactive,zhang2020modeling,sun2021point}, for each user, we chronologically divide his check-in records into different trajectories by day.  
As region is the essential coarse-grained geographical attribute of POIs, we thus group
POIs in each city into 9 geographically-close clusters via k-means, where each POI cluster denotes a region. Fig.~\ref{fig:K-means} visualizes the regions of the four cities. 
To further validate the quality of our clustered regions, for each city, we take the region with the largest number of POIs in Fig.~\ref{fig:K-means} as the city center and find that it exactly coincides with the real CBD area in that city. 
The statistics of the four datasets are summarized in Table~\ref{tab:data_statistic}.

\subsection{Exploring User Behavior Patterns Regarding POI Regions}\label{subsec:explore-behavior}

To better understand user behavior patterns regarding the regions of their visited POIs, we attempt to answer the following five questions.

\medskip\noindent\textbf{Q1.} \textit{How many regions users visited in their daily trajectories and what are these regions?}
To answer this question, we conduct analysis from the perspective of trajectories and users. Fig.~\ref{fig:analysis_results}(a) shows the trajectory distribution w.r.t. the number of visited regions in each trajectory, where we can observe that over 80\% of the trajectories only contain check-ins on one region across the four cities.
From user perspective, we first count the largest number of regions visited by each user among all their daily trajectories, then obtain the user distribution w.r.t the maximum number of daily-visited regions, as shown in Fig.~\ref{fig:analysis_results}(b). We can infer that over 70\% of the users visited at most two regions across all their trajectories, i.e., users tend to be active in no more than two regions rather than all regions. To further dive into what the two regions are, we select trajectories containing no more than two regions for each user, then rank the regions in each trajectory w.r.t individual-user visiting frequency among all the nine regions. Fig.~\ref{fig:analysis_results}(c) depicts the trajectory distribution w.r.t the personalized rank of regions, which indicates that around 60\% of the trajectories contain the user's top \emph{frequently-visited region}, and almost 80\% of the trajectories contain the user's top-2 most \emph{frequently-visited regions}. 
To conclude, most of the users visit only one region in a day and the region is prone to be their most \emph{frequently-visited region}.

\begin{figure}[t]
\centering
\includegraphics[width=0.45\linewidth]{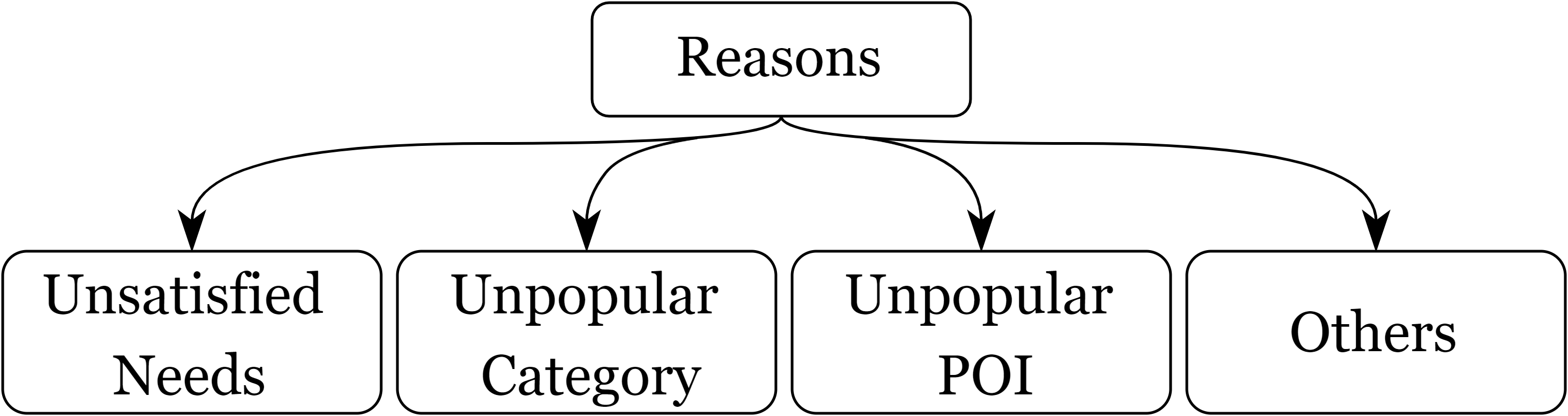}
\caption{The reasons for users visiting infrequently-visited regions or crossing regions.
}\label{fig:illustration_reason}
\end{figure}

\medskip\noindent\textbf{Q2.} \textit{Why do users visit infrequently-visited regions (i.e., not the user's top frequently-visited region)? }
Although users tend to visit POIs in their top frequently-visited regions based on \textbf{Q1}, there are around 20\% of trajectories containing more than one region in Fig.~\ref{fig:analysis_results}(a). To interpret the motivation of users visiting infrequently-visited regions, we compare all the POIs in the frequently- and infrequently-visited regions, and summarize the possible reasons in Fig.~\ref{fig:illustration_reason}. 

Let's take Bob visiting a bar `HopHeads' in his infrequently-visited region `Clarke Quay' as an example. The first reason for his movement may be \textit{`Unsatisfied Needs'}, indicating that
there is no bar in his frequently-visited region, so that his need cannot be met. \textit{`Unpopular Category'} means that the popularity of the category `bar' in his frequently-visited region is lower than that in his infrequently-visited regions `Clarke Quay', which is famous for `bar' in Singapore. 
Note that, the category popularity in a region is the total check-in frequency of all POIs under this category within the region. 
\textit{`Unpopular POI'} denotes that the most popular bar in his frequently-visited region is not as popular as `HopHeads' in his infrequently-visited region. In other words, Bob is likely to be attracted by a specific popular location in his infrequently-visited region. 
Besides, there are also \textit{`Other'} reasons, for example, users may need to move from their frequently-visited region (e.g., office area) to their infrequently-visited region (e.g. private area).

\begin{figure}[t]
\centering
\includegraphics[width=0.95\linewidth]{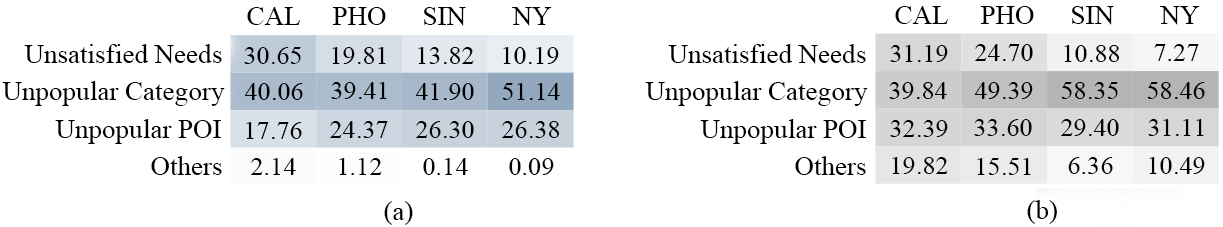}
\caption{The percentage of various reasons for (a) visiting infrequently-visited regions; and (b) crossing regions.
}\label{fig:reason_percentage}
\vspace{-0.1in}
\end{figure}

Fig.~\ref{fig:reason_percentage}(a) shows the percentage of POIs users visited in their infrequently-visited region w.r.t various reasons. On average, around 40\% of POIs in infrequently-visited regions are visited due to \textit{`Unpopular Category'} and 20\% or so owing to \textit{`Unpopular POI'}. 
Besides, a slight difference between the four cities w.r.t the \textit{`Unsatisfied Needs'} can be noted, that is, its percentage in CAL and PHO is much higher than that in NY and SIN. This is mainly because many categories are absent in some regions of CAL and PHO as POIs in the two cities are sparser than the other two cities.

\medskip\noindent\textbf{Q3.} \textit{Why do users choose another region to visit instead of the current region (i.e., cross-region check-in)?} We conduct analysis on all successive cross-region check-in pairs (e.g., $r_t\rightarrow r_{t+1}$, $r_t\neq r_{t+1}$) to analyze the possible reasons, which are similar as those in Fig.~\ref{fig:illustration_reason}. To illustrate, we also take Bob moving from his current region $r_t$ at time $t$ to visit a bar `HopHeads' in another region $r_{t+1}$ at time $t+1$, as an example. 
\textit{`Unsatisfied Needs'} means that there are no bars in region $r_t$, but available in region $r_{t+1}$; \textit{`Unpopular Category'} indicates that the popularity of the category `bar' in $r_t$ is lower than that in $r_{t+1}$; \textit{`Unpopular POI'} suggests that the most trendy bar in $r_t$ is less popular than `HopHeads' in $r_{t+1}$; for \textit{`Others'}, the possible reason is, for example, one user needs to move from his office in $r_t$ to his home in $r_{t+1}$. Fig.~\ref{fig:reason_percentage}(b) displays the percentage of these reasons, where the trend is consistent with Fig.~\ref{fig:reason_percentage}(a): \textit{`Unpopular Category'} is the primary reason for attracting users to cross regions, followed by \textit{`Unpopular POI'}.

\medskip\noindent\textbf{Q4.} \textit{When do users cross regions?} As time plays an vital role in next POI recommendation~\cite{sun2021point}, we explore how time affects users' cross-region behaviors. Specifically, for all successive cross-region check-in pairs (e.g., $r_t\rightarrow r_{t+1}$, $r_t\neq r_{t+1}$), we study the time interval between check-ins in regions $r_t$ and $r_{t+1}$. 
Fig.~\ref{fig:analysis_results}(d) depicts the distribution of cross-region check-in pairs w.r.t the check-in time interval.
It reveals that the time interval between the successive check-ins in $r_{t}$ and $r_{t+1}$ is less than 4 hours for almost 70\% of the cross-region check-in pairs. Alternatively stated,  there is a small probability of the user moving to a new region if the time interval is bigger than 4 hours. Hence, we further split a day into six time slots with each having four hours to investigate the peak period of cross-region behaviors. As shown in Fig.~\ref{fig:analysis_results}(e), the peak period 
is mainly concentrated from 12 pm to 4 pm. That is to say, the probability of a user moving towards another region from the current region
at, e.g., 2 pm, is higher than at, e.g., 9 pm. 

\medskip\noindent\textbf{Q5.} \textit{What factors will affect users to choose which region to visit when they plan to cross region?} We analyze all successive cross-region check-in pairs from three aspects, i.e., distance, category popularity and types of regions (i.e., frequently- or infrequently-visited regions). 

As shown in Fig.~\ref{fig:K-means}, each city has been divided into nine regions. Accordingly, we first calculate the distance between region $r_{t}$ and all the other eight regions via the distance between the centers of these regions, and then check the rank of the distance between $r_{t}$ and $r_{t+1}$. Note that, shorter distance indicates higher rank. Fig.~\ref{fig:analysis_results}(f) displays the distribution of successive cross-region check-in pairs w.r.t the distance rank of $r_t \rightarrow r_{t+1}$. On average, almost 50\% of the next visited regions $r_{t+1}$ are the top-2 closest regions with the current region $r_{t}$.

To explore the effect of category popularity, we first find out all the regions containing category $c_{t+1}$ that is visited by the user in $r_{t+1}$. We then rank these regions by the popularity of category $c_{t+1}$ in them and get the rank of $r_{t+1}$ among them. Note that, higher popularity indicates higher rank. Fig.~\ref{fig:analysis_results}(g) depicts the distribution of successive cross-region check-in pairs w.r.t the rank of regions $r_{t+1}$. We can observe that around 25\% of the next visited regions $r_{t+1}$ are the highest ranked regions, and over 50\% are the top-3 ranked regions. To sum up, users prefer to check-in the next regions $r_{t+1}$ with popular $c_{t+1}$ when they plan to cross region. 

To analyze the impact of different region types (i.e., frequently- or infrequently-visited regions), we consider three types of successive cross-region behaviors: frequently-visited region $\rightarrow$ infrequently-visited region (fre$\rightarrow$inf),  infrequently-visited region $\rightarrow$ frequently-visited region (inf$\rightarrow$fre) and infrequently-visited $\rightarrow$ infrequently-visited region (inf$\rightarrow$inf). 
Fig.~\ref{fig:analysis_results}(h) exhibits the proportion of the three types of cross-region behaviors. We observe that the percentage of both (fre$\rightarrow$inf) and (inf$\rightarrow$fre) are higher than (inf$\rightarrow$inf), indicating that users are more inclined to transit between frequent-visited regions and infrequently-visited regions.

%% file: plot/region_k-means.tex
\begin{figure*}[t]
\centering
\graphicspath{{./figure/}} 
\begin{minipage}{0.45\linewidth}
  \centerline{\includegraphics[width=6cm]{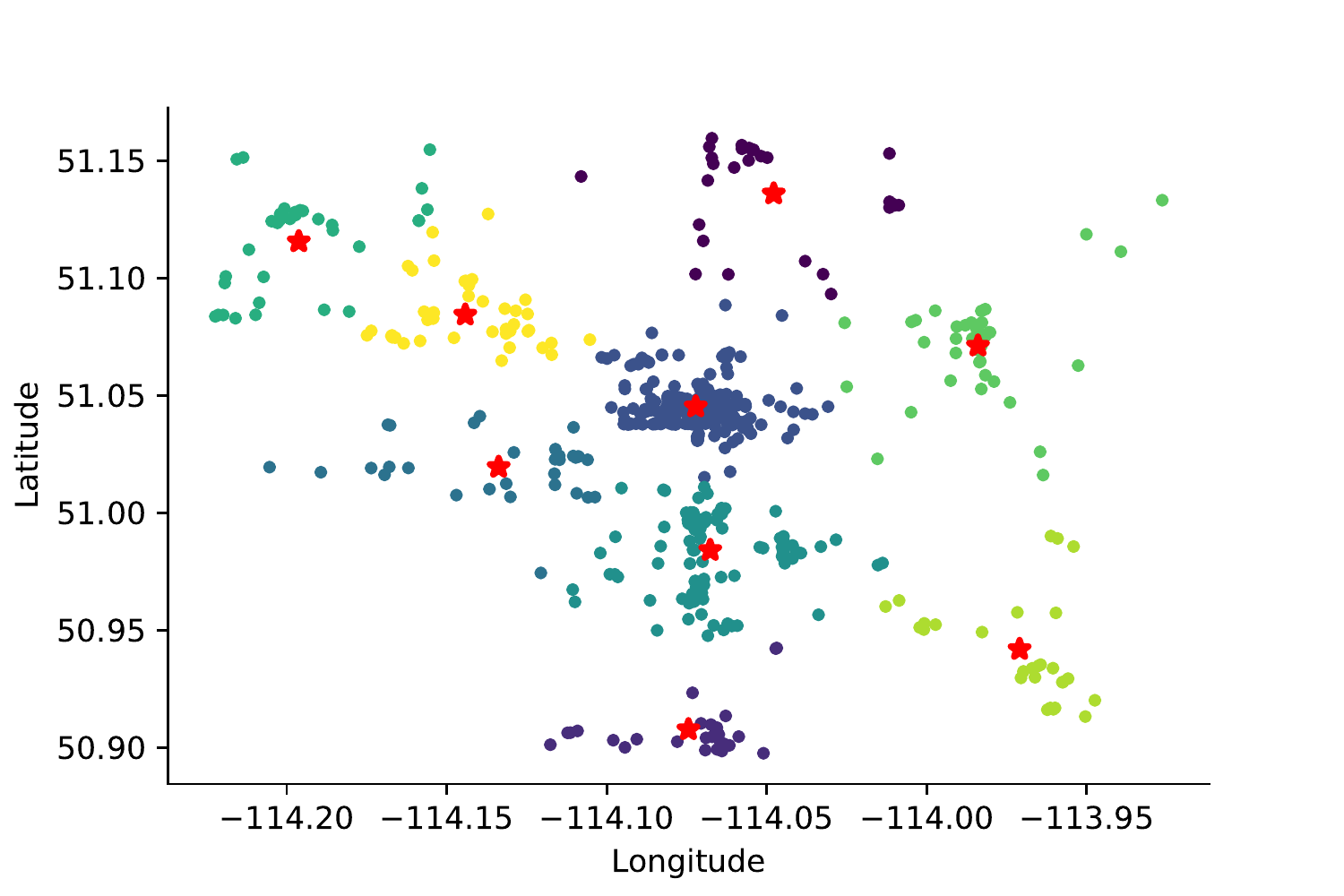}}
  \centerline{\scriptsize{(a) CAL}}
\end{minipage}
\begin{minipage}{0.45\linewidth}
  \centerline{\includegraphics[width=6cm]{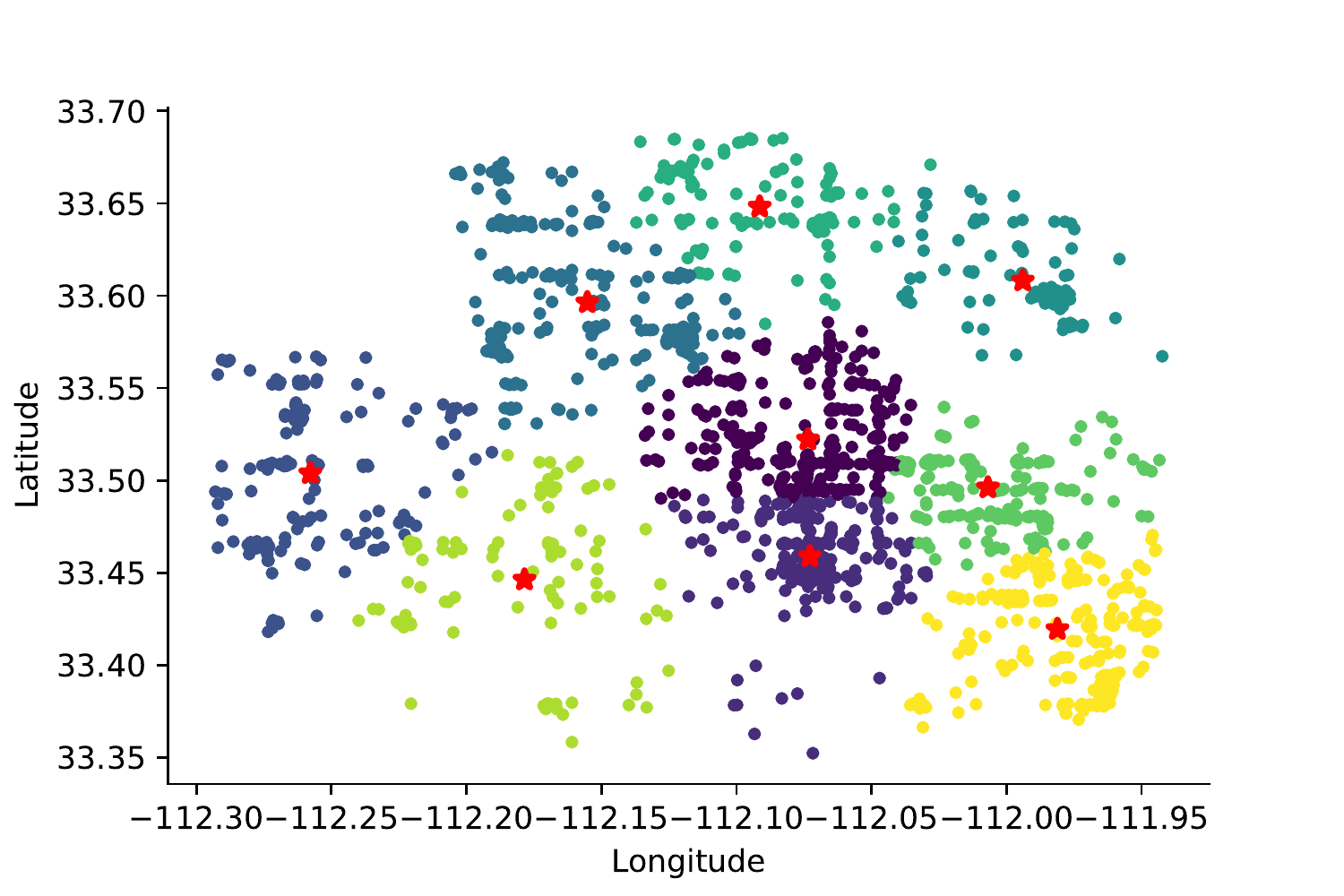}}
  \centerline{\scriptsize{(b) PHO}}
\end{minipage}

\begin{minipage}{0.45\linewidth}
  \centerline{\includegraphics[width=6cm]{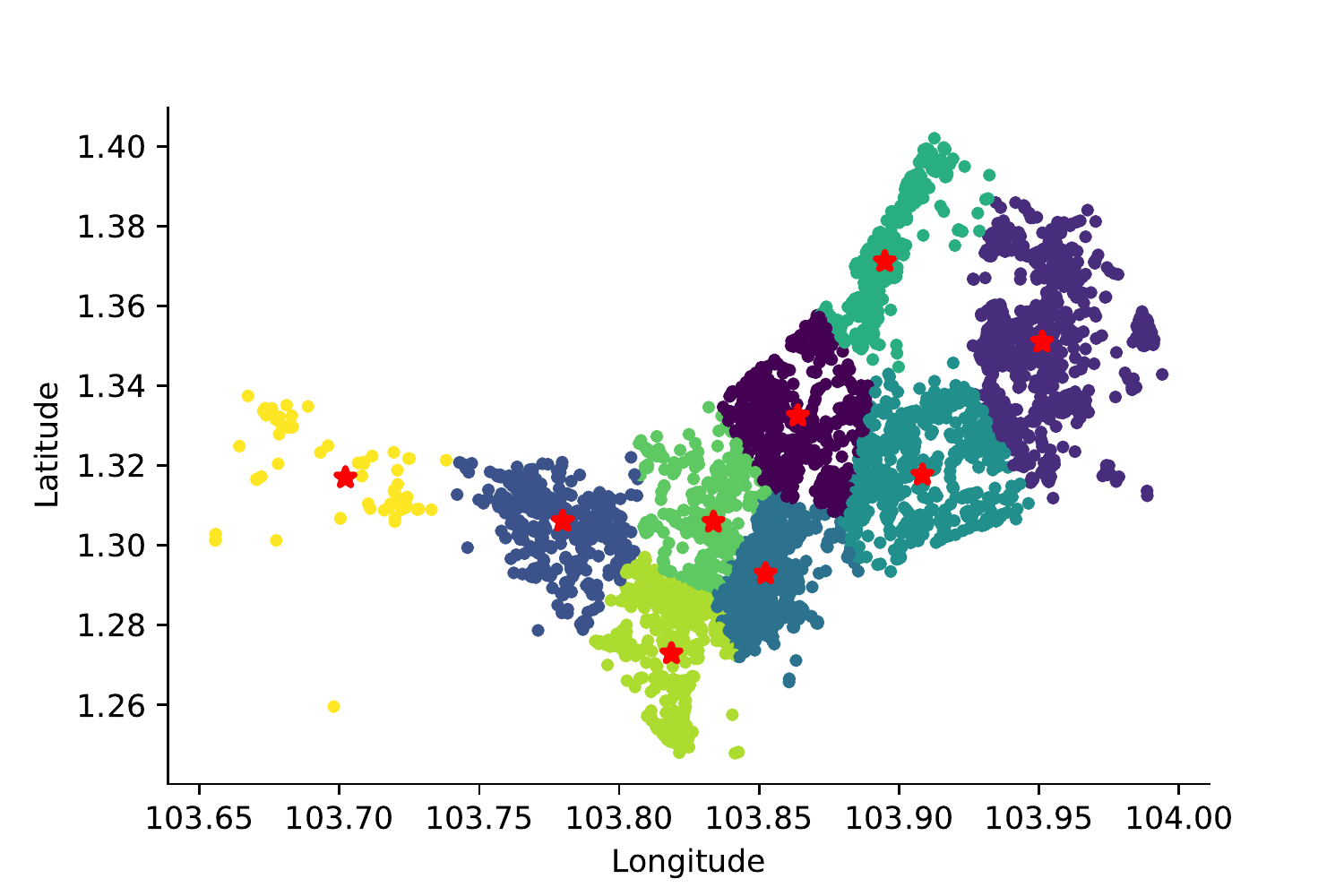}}
  \centerline{\scriptsize{(c) SIN}}
\end{minipage}
\begin{minipage}{0.45\linewidth}
  \centerline{\includegraphics[width=6cm]{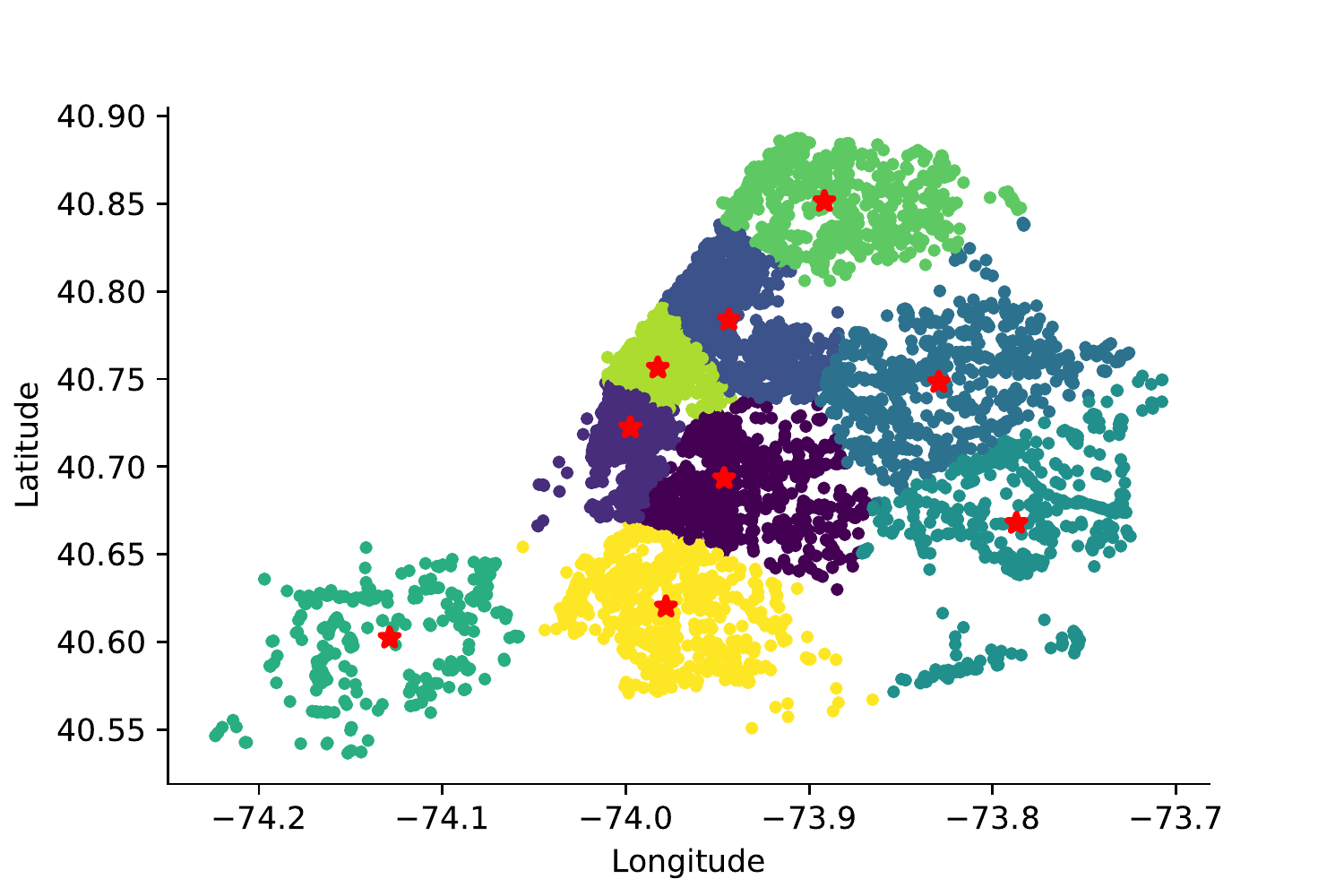}}
  \centerline{\scriptsize{(d) NY}}
\end{minipage}
\caption{The regions clustered via k-means for the four cities, where the dots with the same color denote a region; and the red star within each region represent the center of that region, calculated by averaging the geographical positions of all POIs in that region.}
\label{fig:K-means}
\vspace{-0.1in}
\end{figure*}


%% file: table/statistics.tex
\begin{table}[t]
\renewcommand\arraystretch{1.2}
\caption{Statistics of the four real-world datasets utilized in our study.}\label{tab:data_statistic}
\vspace{-0.1in}
  \begin{tabular}{l|rrrccc}
    \specialrule{.15em}{.15em}{.15em}
    Dataset&\#Users&\#POIs &\#Check-ins &\#Regions&\#Categories &Density\\
    \specialrule{.1em}{.1em}{.1em}
    CAL&435&3,013 &13,911 &9 &293 & 1.06\%\\
    PHO&2,946&7,247 &47,980 &9&344 & 0.22\%\\
    SIN&8,648 &33,712 &355,337 &9 &398 & 0.12\%\\
    NY&16,387&56,252 &511,431 &9&420 & 0.06\%\\
  \specialrule{.15em}{.15em}{.15em}
\end{tabular}
\vspace{-0.1in}
\end{table}

%% file: plot/analysis_results.tex
\pgfplotstableread{ 
Label i2i i2a a2i
CAL   26.54 33.33 40.13
PHO   27.03 36.75 36.22
SIN   28.81 35.32 35.87
NY    27.78 35.44 36.78
}\plotdata

\begin{figure*}[ht]
	\setlength{\belowcaptionskip}{-0.12in}
	\centering
	\subfigure[\scriptsize Trajectory distribution]{
	\begin{tikzpicture}[xscale=0.56, yscale=0.49]
	\begin{axis}[
    ybar, 
    bar width=2pt,
    enlarge x limits= 0.08, 
    symbolic x  coords = {1,2,3,4,5,6,7,8,9},
    xtick={1,2,3,4,5,6,7,8,9},
    xlabel = {\# visited regions},
    xlabel style ={font = \Large},
    ymin=0, ymax=0.9, 
    ylabel = {ratio of trajectories},
    ytick={0,0.2,0.4,0.6,0.8,1},
    ylabel style ={font = \Large},
	tick label style={font = \Large},
    yticklabel style = {/pgf/number format/.cd,fixed,precision=3},
    ymajorgrids=true,
    grid style=dashed,
	legend style={at={(0.8,0.8)},font = \Large, anchor=north,legend columns=1,row sep=0.2cm,legend cell align={left},draw=none},
	ymajorgrids=true,
	grid style=dashed,
	]
	\addplot[color=4,fill=4] coordinates {
		(1, 0.8304134548002803)
    	(2, 0.14838822704975474)
    	(3, 0.01857042747021724)
    	(4, 0.0026278906797477224)
        (5,0.0)
        (6,0.0)
        (7,0.0)
        (8,0.0)
        (9,0.0)
	};			
	\addplot[color=9,fill=9] coordinates {
		(1, 0.8719721715521995)
    	(2, 0.11267127646035718)
    	(3, 0.014338438043524372)
    	(4, 0.000721164043609214)
    	(5, 0.00025452848597972257)
    	(6, 4.242141432995376e-05)
        (7,0.0)
        (8,0.0)
        (9,0.0)
	};			
	\addplot[color=7,fill=7] coordinates {
		(1, 0.8222836095764272)       (2, 0.1520670603924557)       (3, 0.02214390042547787)       (4, 0.0029656442496983552)       (5, 0.000444529116657141)       (6, 6.350415952244872e-05)       (7, 3.175207976122436e-05)
        (8,0.0)
        (9,0.0)
	};	
	\addplot[color=6,fill=6] coordinates {
		(1, 0.837756126318473)
    	(2, 0.13998916541856538)
    	(3, 0.018546254102801057)
    	(4, 0.002696695452662439)
    	(5, 0.0006851279436601766)
    	(6, 0.00027086453586565117)
    	(7, 5.5766227972339954e-05)
        (8,0.0)
        (9,0.0)
	};
	\legend{CAL, PHO, SIN, NY}
	\end{axis}
	\end{tikzpicture}}\hspace{0.5in}
	\subfigure[\scriptsize User distribution]{
	\begin{tikzpicture}[xscale=0.56, yscale=0.49]
	\begin{axis}[
    ybar, 
    bar width=2pt,
    enlarge x limits= 0.08, 
    symbolic x  coords = {1,2,3,4,5,6,7,8,9},
        xtick={1,2,3,4,5,6,7,8,9},
        xlabel = {\# visited regions},
        xlabel style ={font = \Large},
        ymin=0, ymax=0.65, 
        ylabel = {ratio of users},
        ytick={0,0.2,0.4,0.6},
        ylabel style ={font = \Large},
		tick label style={font = \Large},
        yticklabel style = {/pgf/number format/.cd,fixed,precision=3},
        ymajorgrids=true,
	    grid style=dashed,
	legend style={at={(0.8,0.8)},font = \Large, anchor=north,legend columns=1,row sep=0.2cm,legend cell align={left},draw=none},
	ymajorgrids=true,
	grid style=dashed,
	]
	\addplot[color=4,fill=4] coordinates {
		(1, 0.36153846153846153)       (2, 0.49230769230769234)       (3, 0.1076923076923077)       (4, 0.038461538461538464)       (5, 0.0)       (6, 0.0)       (7, 0.0)       (8, 0.0)   (9,0.0)
	};			
	\addplot[color=9,fill=9] coordinates {
		(1, 0.4002607561929596)       (2, 0.4445893089960887)       (3, 0.1408083441981747)       (4, 0.009126466753585397)       (5, 0.003911342894393742)       (6, 0.001303780964797914)       (7, 0.0)       (8, 0.0)  (9, 0.0)
	};			
	\addplot[color=7,fill=7] coordinates {
		(1, 0.18909012505390255)       (2, 0.5217766278568349)       (3, 0.23803363518758086)       (4, 0.041828374299266925)       (5, 0.007330746011211729)       (6, 0.00129366106080207)       (7, 0.000646830530401035)       (8, 0.0)  (9,0.0)
	};	
	\addplot[color=6,fill=6] coordinates {
		(1, 0.20255165405893644)       (2, 0.5710737269955967)       (3, 0.17545444281359376)       (4, 0.03353279891611155)       (5, 0.010500169357570283)       (6, 0.005419442249068533)       (7, 0.0014677656091227277)       (8, 0.0)   (9, 0.0)
	};
	\legend{CAL, PHO, SIN, NY}
	\end{axis}
	\end{tikzpicture}
	}
	
	\subfigure[\scriptsize Trajectory distribution]{
	\begin{tikzpicture}[xscale=0.56, yscale=0.49]
	\begin{axis}[
    ybar, 
    bar width=2pt,
    enlarge x limits= 0.08, 
    symbolic x  coords = {1,2,3,4,5,6,7,8,9},
    xtick={1,2,3,4,5,6,7,8,9},
    xlabel = {region ranking w.r.t visiting frequency},
    xlabel style ={font = \Large},
    ymin=0, ymax=0.7, 
    ylabel = {ratio of trajectories},
    ytick={0,0.2,0.4,0.6,0.8},
    ylabel style ={font = \Large},
	tick label style={font = \Large},
    yticklabel style = {/pgf/number format/.cd,fixed,precision=3},
    ymajorgrids=true,
    grid style=dashed,
	legend style={at={(0.8,0.8)},font = \Large, anchor=north,legend columns=1,row sep=0.2cm,legend cell align={left},draw=none},
	ymajorgrids=true,
	grid style=dashed,
	]
	\addplot[color=4,fill=4] coordinates {
		(1,0.6156356854211999)
        (2,0.2326701896176562)
        (3,0.09278831209201119)
        (4,0.032639104755983834)
        (5,0.015076157911097297)
        (6,0.00699409387628225)
        (7,0.003730183400683867)
        (8,0.00046627292508548336)
        (9,0.0)
	};			
	\addplot[color=9,fill=9] coordinates {
		(1,0.6266285228283141)
        (2,0.21251014806510227)
        (3,0.08667414079715467)
        (4,0.04082421618278115)
        (5,0.018943054857540496)
        (6,0.009239571655004446)
        (7,0.0040205667452739013)
        (8,0.000927823095063208)
        (9,0.000231955773765802)
	};			
	\addplot[color=7,fill=7] coordinates {
		(1,0.55728194749037364)
        (2,0.22700801118521568)
        (3,0.10351962204795436)
        (4,0.05252653951752482)
        (5,0.02988549811982388)
        (6,0.017437435518697464)
        (7,0.009223292760617217)
        (8,0.0028752318508036556)
        (9,0.00024242150898932783)
	};		
	\addplot[color=6,fill=6] coordinates {
		(1,0.5779975410274229)
        (2,0.2551593877514656)
        (3,0.09835177562766166)
        (4,0.039646478145435755)
        (5,0.01764758290123127)
        (6,0.00755510414996169)
        (7,0.002822472871117763)
        (8,0.000698490761034194)
        (9,0.00012116676466919692)
	};
	\legend{CAL, PHO, SIN, NY}
	\end{axis}
	\end{tikzpicture}
	}\hspace{0.5in}
	\subfigure[\scriptsize Distribution w.r.t. time interval]{
        \begin{tikzpicture}[xscale=0.56, yscale=0.53, font=\Large, thick, /pgfplots/boxplot/box extend=0.4]
          \begin{axis}
            [
            xtick={1,2,3,4},
            ylabel={time interval},
            ylabel style ={font = \Large},
            xticklabels={CAL, PHO, SIN, NY},
            ytick={1,4,7,10,13,16,19,22},
            boxplot/draw direction=y,
            xlabel style ={font = \Large},
            tick label style={font = \Large},
            ymajorgrids=true,
        	grid style=dashed,
        	enlarge y limits= 0.05,
            ]
            \addplot+[
            color = black,
            fill = 4,
            boxplot prepared={
              median=2,
              upper quartile=4,
              lower quartile=1,
              upper whisker=24,
              lower whisker=1
            },
            ] coordinates {};
            \addplot+[
            color = black,
            fill = 9,
            boxplot prepared={
              median=2,
              upper quartile=4,
              lower quartile=1,
              upper whisker=24,
              lower whisker=1,
            },
            ] coordinates {};
            \addplot+[
            color = black,
            fill = 7,
            boxplot prepared={
              median=2,
              upper quartile=4.5,
              lower quartile=2,
              upper whisker=24,
              lower whisker=1
            },
            ] coordinates {};
            \addplot+[
            color = black,
            fill = 6,
            boxplot prepared={
              median=3,
              upper quartile=4,
              lower quartile=2,
              upper whisker=24,
              lower whisker=1
            },
            ] coordinates {};
          \end{axis}
        \end{tikzpicture}}
	
	\subfigure[\scriptsize Distribution w.r.t. time slots]{
		\begin{tikzpicture}[xscale=0.56, yscale=0.49]
	\begin{axis}[
    ybar, 
    bar width=2pt,
    xtick={4,8,12,16,20,24},
    xlabel style ={font = \Large},
    xticklabels={0-4,4-8,8-12,12-16,16-20,20-24},
    ymin=0, ymax=37,
    ylabel = {ratio of cross-region pairs},
    ytick={0,10,20,30},
    ylabel style ={font = \Large},
	tick label style={font = \Large},
    yticklabel style = {/pgf/number format/.cd,fixed,precision=3},
    ymajorgrids=true,
    grid style=dashed,
	legend style={at={(0.25,0.96)},font = \Large, anchor=north,legend columns=2,row sep=0.2cm,legend cell align={left},draw=none},
	ymajorgrids=true,
	grid style=dashed,
	]
	\addplot[color=4,fill=4] coordinates {
		(4, 6.1368209255533195)        (8, 20.72434607645875)        (12, 21.730382293762574)        (16, 34.50704225352113)        (20, 23.239436619718308)        (24, 6.237424547283702)
	};			
	\addplot[color=9,fill=9] coordinates {
		(4, 7.55420054200542)        (8, 17.378048780487806)        (12, 25.101626016260163)        (16, 30.521680216802167)        (20, 26.761517615176153)        (24, 12.940379403794038)
	};			
	\addplot[color=7,fill=7] coordinates {
		((4, 8.798862310385065)        (8, 16.379813302217038)        (12, 23.588827304550758)        (16, 33.64935822637106)        (20, 30.593640606767796)        (24, 16.92313302217036)
	};
	\addplot[color=6,fill=6] coordinates {
		(4, 11.50542447300552)        (8, 15.663831873308183)        (12, 22.98527186307735)        (16, 26.649188993328643)        (20, 25.457723213334187)        (24, 22.102862501864994) 
	};
	\legend{CAL, PHO, SIN, NY}
	\end{axis}
	\end{tikzpicture}
	}\hspace{0.5in}
	\subfigure[\scriptsize Distribution w.r.t. distance ]{
        \begin{tikzpicture}[xscale=0.56, yscale=0.49, thick, /pgfplots/boxplot/box extend=0.4]
          \begin{axis}
            [
            xtick={1,2,3,4},
            xticklabels={CAL, PHO, SIN, NY},
            ytick={1,2,3,4,5,6,7,8},
            ylabel={distance rank},
            ylabel style ={font = \Large},
            boxplot/draw direction=y,
            ylabel style ={font = \Large},
           tick label style={font = \Large},
           ymajorgrids=true,
        	grid style=dashed,
            ]
            \addplot+[
            color = black,
            fill = 4,
            boxplot prepared={
              median=2.5,
              upper quartile=4,
              lower quartile=1.5,
              upper whisker=8,
              lower whisker=1
            },
            ] coordinates {};
            \addplot+[
            color = black,
            fill = 9,
            boxplot prepared={
              median=2,
              upper quartile=3,
              lower quartile=1,
              upper whisker=8,
              lower whisker=1,
            },
            ] coordinates {};
            \addplot+[
            color = black,
            fill = 7,
            boxplot prepared={
              median=2,
              upper quartile=5,
              lower quartile=1,
              upper whisker=8,
              lower whisker=1
            },
            ] coordinates {};
            \addplot+[
            color = black,
            fill = 6,
            boxplot prepared={
              median=2,
              upper quartile=3,
              lower quartile=1,
              upper whisker=8,
              lower whisker=1
            },
            ] coordinates {};
          \end{axis}
        \end{tikzpicture}}
        
    \subfigure[\scriptsize Distribution w.r.t. category popularity]{
        \begin{tikzpicture}[xscale=0.56, yscale=0.49, thick, /pgfplots/boxplot/box extend=0.4]
          \begin{axis}
            [
            xtick={1,2,3,4},
            xticklabels={CAL, PHO, SIN, NY},
            ytick={1,2,3,4,5,6,7,8,9},
            ylabel={region rank},
            ylabel style ={font = \Large},
            boxplot/draw direction=y,
            ylabel style ={font = \Large},
            tick label style={font = \Large},
            ymajorgrids=true,
        	grid style=dashed,
            ]
            \addplot+[
            color = black,
            fill = 4,
            boxplot prepared={
              median=2,
              upper quartile=4,
              lower quartile=1,
              upper whisker=9,
              lower whisker=1
            },
            ] coordinates {};
            \addplot+[
            color = black,
            fill = 9,
            boxplot prepared={
              median=2,
              upper quartile=5,
              lower quartile=1,
              upper whisker=9,
              lower whisker=1,
            },
            ] coordinates {};
            \addplot+[
            color = black,
            fill = 7,
            boxplot prepared={
              median=3,
              upper quartile=5,
              lower quartile=1,
              upper whisker=9,
              lower whisker=1
            },
            ] coordinates {};
            \addplot+[
            color = black,
            fill = 6,
            boxplot prepared={
              median=2,
              upper quartile=4,
              lower quartile=1,
              upper whisker=9,
              lower whisker=1
            },
            ] coordinates {};
          \end{axis}
        \end{tikzpicture}}\hspace{0.5in}
    \subfigure[\scriptsize Distribution w.r.t. various transitions]{
    \begin{tikzpicture}[xscale=0.56, yscale=0.49]
        \begin{axis}[
                    ybar stacked,   
                    ymin=0, 
                    ymax=120,
                    ytick={0,50,100},
                    ylabel = {ratio of cross-region pairs},
                    xtick=data,     
                    xticklabels from table={\plotdata}{Label}, 
                    ylabel style ={font = \Large},
	               tick label style={font = \Large},
                    legend style={at={(0.5,0.97)},font = \Large, anchor=north,legend columns=3,legend cell align={left},draw=none},
                	ymajorgrids=true,
                	grid style=dashed,
        ]
        \addplot [color=7,fill=7] table [y=i2i, meta=Label,x expr=\coordindex] {\plotdata};   
        \addplot [color=6,fill=6] table [y=i2a, meta=Label,x expr=\coordindex] {\plotdata};
        \addplot [color=9,fill=9] table [y=a2i, meta=Label,x expr=\coordindex] {\plotdata};
        \legend{ inf$\rightarrow$ inf, inf$\rightarrow$fre, fre$\rightarrow$inf}
        \end{axis}
    \end{tikzpicture}
    }
	\caption{(a-b) trajectory and user distribution w.r.t the number of daily-visited regions; (c) trajectory distribution w.r.t the personalized visiting frequency rank of regions; (d-e) successive cross-region check-in distribution w.r.t different time interval and time slots of a day; (f-g) successive cross-region check-in distribution w.r.t distance and category popularity of various regions; (h) successive cross-region check-in distribution w.r.t the transition of different types of regions.
	}\label{fig:analysis_results}
 \vspace{-0.1in}
\end{figure*}

%% file: section/method.tex
\section{The Proposed MCMG}

\begin{figure}[t]
\centering
\includegraphics[width=0.9\linewidth]{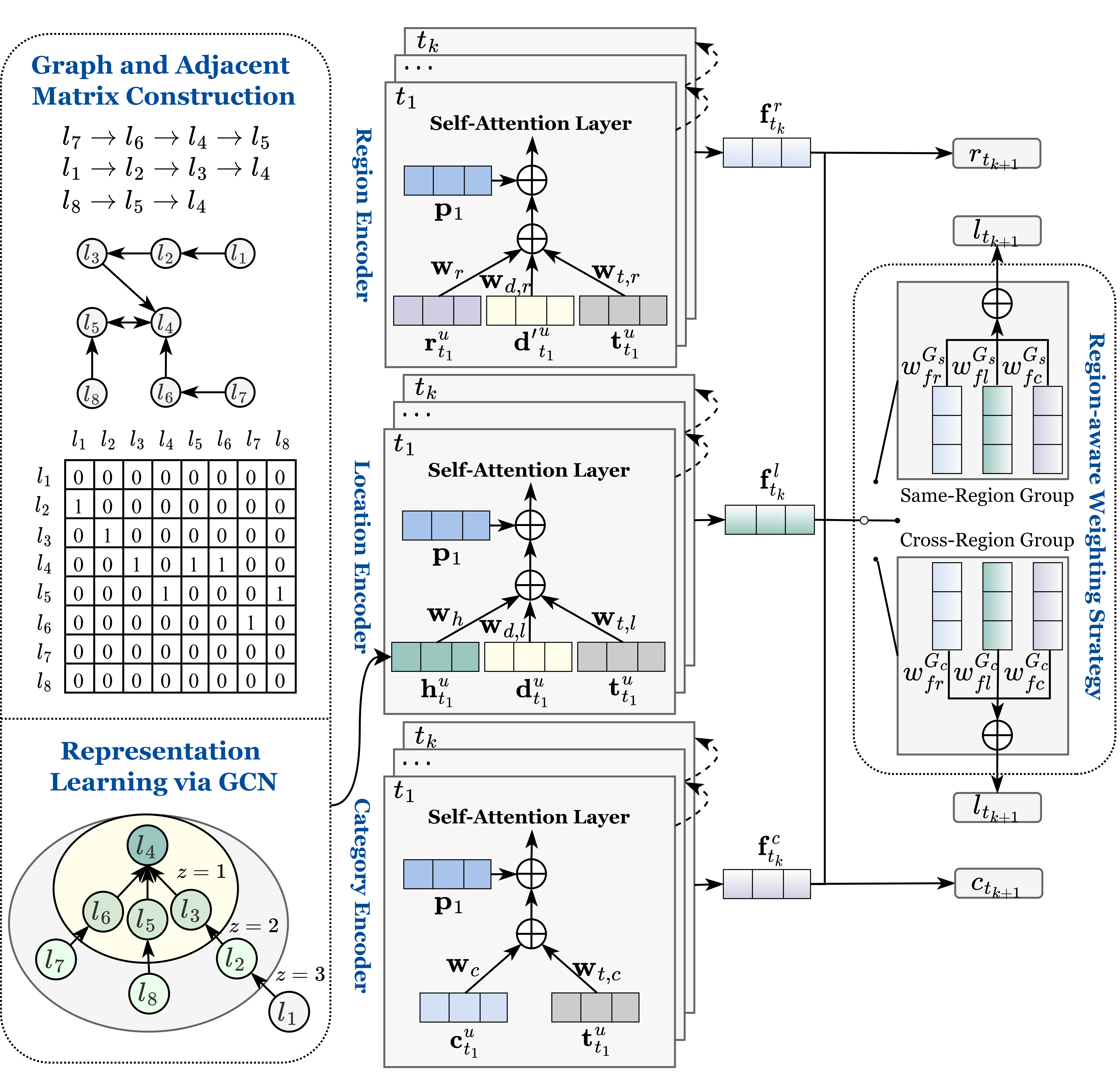}
\caption{The overall architecture of MCMG. 
It takes POI $l_4$ to illustrate the representation learning via GCN. 
}\label{fig:model_framework}
\vspace{-0.1in}
\end{figure}

In this section, we introduce the proposed \underline{M}ulti-\underline{C}hannel next POI recommendation framework with \underline{M}ulti-\underline{G}ranularity check-in signals (MCMG) from the two orthogonal perspectives. The overall architecture of MCMG is presented in Fig.~\ref{fig:information_granularity}.

\subsection{Preliminaries and Model Overview}
\textbf{Notations.} 
Let $\mathcal{U}=\{u_1,u_2,\cdots,u_{|\mathcal{U}|}\}$, $\mathcal{L}=\{l_1,l_2,\cdots,l_{|\mathcal{L}|}\}$, $\mathcal{R}=\{r_1,r_2,\cdots,r_9\}$ and $\mathcal{C}=\{c_1,c_2,\cdots,c_{|\mathcal{C}|}\}$ denote the sets of users, POIs (locations), regions and categories of POIs, respectively. A check-in record $(u,l,t,r,c,g)$ indicates user $u$ visited POI $l$ at time $t$, where $l$ is characterized by region $r$, category $c$ and geocoded by $g$ (longitude, latitude). Each region is geocoded by $g'$, i.e., the longitude and latitude of region center. For each user, we chronologically order all his historical check-in records, and then split them into trajectories by day~\cite{zhang2020interactive}. 
The $i$-th POI trajectory of user $u$ is denoted as $\mathcal{L}^{u,i}=\{L_{t_1}^u, L_{t_2}^u,\cdots, L_{t_k}^u\}$, and $L_{t_k}^u = (l_{t_k}^u, t_k^u, g_{t_k}^u)$; the corresponding region and category check-in trajectories are respectively denoted as $\mathcal{R}^{u,i}=\{R_{t_1}^u, R_{t_2}^u,\cdots, R_{t_k}^u\}$ and $\mathcal{C}^{u,i}=\{C_{t_1}^u, C_{t_2}^u,\cdots, C_{t_k}^u\}$, where $R_{t_k}^u = (r_{t_k}^u, t_k^u, {g'}_{t_k}^u)$ and $C_{t_k}^u = (c_{t_k}^u, t_k^u)$.
Given $\mathcal{L}^{u,i}$, $\mathcal{R}^{u,i}$, and $\mathcal{C}^{u,i}$, our goal is to predict POI $l_{t_{k+1}}$ for user $u$ at the next time $t_{k+1}$. Important notations used in this paper are summarized in Table~\ref{table:notations}.

\input{table/notation.tex}

\medskip\noindent\textbf{Model Overview.}
Fig.~\ref{fig:model_framework} depicts the overall architecture of MCMG, which mainly composed of three modules. (1) \textit{Global user behavior encoder} aims to capture the augmented sequential regularity in global all-user POI check-ins for more expressive POI representation learning. 
(2) \textit{Local multi-channel encoder} equips with location, region and category encoders to learn the transition patterns w.r.t. POI, region and activity from the local individual-user check-ins. (3) \textit{Region-aware weighting strategy} seeks to dynamically balance the impact of different local channels, thus avoiding potential contradictory predictions and achieving more accurate next POI recommendation. We will elaborate each module in what follows.

\subsection{Global User Behavior Encoder}
The purpose of this module is to capture the augmented sequential regularity from the crowd, i.e., global all-user POI check-ins (\textcircled{3} in Fig.~\ref{fig:information_granularity}) for more expressive POI representation learning.  

\medskip\noindent\textbf{Directed POI Graph Construction.}
We first construct a directed POI graph based on all-user POI check-in trajectories. 
Given a POI check-in trajectory, each POI inside is treated as a node, and each successive visited POI pair (e.g., $l_{t_{k-1}}$ and $l_{t_k}$) is linked via a directed edge. By traversing all-user trajectories, the constructed graph can help encode the global-level sequential regularity, so as to assist the local-level user preference learning. The graph is then converted as an in-degree adjacent matrix $\mathbf{A}$ showing the connections of incoming edges. The top left part of Fig.~\ref{fig:model_framework} illustrates a toy example for constructing the corresponding directed graph and in-degree adjacent matrix for the three given trajectories.


\medskip\noindent\textbf{Representation Learning via GCN.} Given the constructed directed POI graph, we adopt GCN~\cite{kipf2016semi,chen2020revisiting} to learn more expressive POI embeddings\footnote{The two terms, i.e., representation and embedding, are exchangeable in this paper.} thanks to its superior capability on representation learning with the graph data, 
\begin{equation}
    \mathbf{H}^{(z+1)}=ReLu(\widetilde{\mathbf{D}}^{-1}\widetilde{\mathbf{A}}\mathbf{H}^{(z)}\mathbf{W}^{(z)}),
\end{equation}
where $\mathbf{A}\in\mathbb{R}^{\vert \mathcal{L}\vert\times\vert \mathcal{L}\vert}$ is the in-degree adjacent matrix; $\widetilde{\mathbf{A}}= \mathbf{A}+\mathbf{I}$; $\mathbf{I}$ is the identity matrix showing the self-connection of each node; $\widetilde{\mathbf{D}}\in\mathbb{R}^{\vert \mathcal{L}\vert\times\vert \mathcal{L}\vert}$ is the diagonal in-degree matrix with $\widetilde{\mathbf{D}}_{ii}=\sum\nolimits_j\widetilde{\mathbf{A}}_{ij}$; $\mathbf{H}^{(z)}\in \mathbb{R}^{\vert\mathcal{L}\vert\times d}$ is the POI embedding matrix in the $z$-th layer; $d$ is the embedding size; $\mathbf{H}^{(0)}$ is the initialized POI embedding matrix; $\mathbf{W}^{(z)}\in \mathbb{R}^{d\times d}$ is a layer-wise trainable weight matrix; and ReLU is the activation function.

\subsection{Local Multi-Channel Encoder}
This module seeks to learn personalized transition patterns w.r.t. location, region and activity from the local individual-user check-ins. It mainly consists of the main channel (location encoder) and two auxiliary channels (i.e., region and category encoders).  

\medskip\noindent\textbf{Location Encoder.}
As our goal is to provide personalized next location prediction for users, it is vital to learn personalized sequential patterns from the local individual-user POI check-ins. To this end, we employ the bidirectional self-attention~\cite{vaswani2017attention} to help 
model each trajectory. This is mainly attributed to (1) the bidirectional self-attention can model both left and right contexts for a target POI, thus can learn more expressive embeddings for POIs; and (2)
it shows superiority in capturing context-aware correlations between non-consecutive check-ins and aggregating the most relevant behaviors automatically within a trajectory~\cite{lin2021pre}. 

Given the global user behavior encoder, the embeddings of all POIs within a trajectory $\mathcal{L}^{u, i}$ can be represented as $\mathbf{H}^{u, i} = [\mathbf{h}^u_{t_1}, \mathbf{h}^u_{t_2}, \cdots, \mathbf{h}^u_{t_k}]$, 
where each hidden state $\mathbf{h}\in\mathbb{R}^d$ is the learned POI embedding in the last layer of GCN. To distinguish the position of different POIs in the trajectory, we further add the position embeddings~\cite{kang2018self} for each POI. Moreover, as user behaviors are highly affected by spatial and temporal contexts~\cite{zhang2020interactive,sun2021point}, we also incorporate these two factors to enrich the POI embeddings. Hence, the enhanced POI embeddings are as follows, 
\begin{equation}
    {\mathbf{\widetilde{H}}}^{u,i} = \begin{bmatrix}   \mathbf{h}^u_{t_1} \mathbf{W}_h+\mathbf{d}^u_1 \mathbf{W}_{d,l}+\mathbf{t}^u_1 \mathbf{W}_{t,l}+\mathbf{p}_1 \\ 
    \mathbf{h}^u_{t_2} \mathbf{W}_h+\mathbf{d}^u_2 \mathbf{W}_{d,l}+\mathbf{t}^u_2 \mathbf{W}_{t,l}+\mathbf{p}_2 \\ \cdots\\
    \mathbf{h}^u_{t_k} \mathbf{W}_h+\mathbf{d}^u_k \mathbf{W}_{d,l}+\mathbf{t}^u_k \mathbf{W}_{t,l}+\mathbf{p}_k  \end{bmatrix},
\end{equation}
where $\mathbf{W}$ is the learnable weight matrix; $\mathbf{d}^u_k\in\mathbb{R}^d$ is the embedding of distance $d^u_k$ between $l_{t_{k-1}}^u$ and $l_{t_{k}}^u$ calculated via $g_{t_{k-1}}^u$ and $g_{t_{k}}^u$; $d_1^u=0$; $\mathbf{t}^u_k\in\mathbb{R}^d$ is the embedding of temporal context by mapping one day into 24 hours; and $\mathbf{p}_k\in\mathbb{R}^d$ is a learnable position embedding.

Subsequently, the enhanced POI embeddings ${\mathbf{\widetilde{H}}}^{u,i}$ is thus fed into the self-attention layer to capture the local sequential correlations, given by,
\begin{equation}\label{equ:SA_1}
    \mathbf{S}^{u,i}_l=softmax( \frac{({\mathbf{\widetilde{H}}}^{u,i}\mathbf{W}^Q)({\mathbf{\widetilde{H}}}^{u,i}\mathbf{W}^K)^T}{\sqrt{d}} )({\mathbf{\widetilde{H}}}^{u,i}\mathbf{W}^V),
\end{equation}
where $\mathbf{S}^{u,i}_l\in\mathbb{R}^{k\times d}$ is the refined embeddings of POIs in $\mathcal{L}^{u,i}$ through the self-attention network; 
$\mathbf{W}^Q, \mathbf{W}^K, \mathbf{W}^V\in\mathbb{R}^{d\times d}$ are the query, key and value projection matrices; and the scale factor $\sqrt{d}$ helps avoid extremely large values of the inner product.
To endow the model with nonlinearity and consider interactions between different latent dimensions~\cite{kang2018self}, a point-wise two-layer feed-forward network is applied to $\mathbf{S}^{u,i}_l$, shown as, 
\begin{equation}\label{equ:SA_2}
    \mathbf{F}^{u,i}_l=ReLU(\mathbf{S}^{u,i}_l \mathbf{W}_1+\mathbf{b}_1)\mathbf{W}_2+\mathbf{b}_2,
\end{equation}
where $\mathbf{F}^{u,i}_l$ is the refined embeddings of POIs in $\mathcal{L}^{u,i}$ through the feed-forward network; and $\mathbf{b}$ is the bias vector. 
To learn more complex location transition patterns and user preference, we can also stack self-attention block with multiple layers~\cite{kang2018self}. 

\medskip\noindent\textbf{Region Encoder.}
Inspired by the observations that user behaviors exhibit noticeable patterns w.r.t. regions of visited POIs (\textbf{Q1}), and user preference on regions are influenced by both temporal and spatial contexts (\textbf{Q4} and \textbf{Q5}) in Section~\ref{sec:data_and_analysis}, the auxiliary region encoder further models region transition patterns by incorporating spatial and temporal contexts. As a result, the enhanced embeddings of regions in the trajectory $\mathcal{R}^{u,i}$ are formulated as below,
\begin{equation}
    \mathbf{R}^{u,i} = \begin{bmatrix}   \mathbf{r}^u_{t_1} \mathbf{W}_r+{\mathbf{d}'}^u_1 \mathbf{W}_{d,r}+\mathbf{t}^u_1 \mathbf{W}_{t,r}+\mathbf{p}_1 \\ 
    \mathbf{r}^u_{t_2} \mathbf{W}_r+{\mathbf{d}'}^u_2 \mathbf{W}_{d,r}+\mathbf{t}^u_2 \mathbf{W}_{t,r}+\mathbf{p}_2 \\ \cdots\\
    \mathbf{r}^u_{t_k} \mathbf{W}_h+{\mathbf{d}'}^u_k \mathbf{W}_{d,r}+\mathbf{t}^u_k \mathbf{W}_{t,r}+\mathbf{p}_k  \end{bmatrix},
\end{equation}
where $\mathbf{r}^u_{t_k}\in \mathbf{R}$ is the embedding of ${r}^u_{t_k}$ in $\mathcal{R}^{u,i}$; $\mathbf{R} \in \mathbb{R}^{9\times d}$ is the region embedding matrix; ${\mathbf{d}'}^u_k\in \mathbb{R}^{d}$ is the embedding of distance ${d'}^u_k$ between $r_{t_{k-1}}^u$ and $r_{t_{k}}^u$ based on ${g'}_{t_{k-1}}^u$ and ${g'}_{t_{k}}^u$; and ${d'}^u_1=0$. 
Afterwards, we then feed $\mathbf{R}^{u,i}$ into the self-attention and two-layer feed-forward networks, which is similar with Eqs. (\ref{equ:SA_1}-\ref{equ:SA_2}). Finally, we can obtain $\mathbf{F}^{u,i}_r$, i.e., the refined embedings of regions in $\mathcal{R}^{u,i}$.

\smallskip\noindent\textbf{Category Encoder.}
The interplay between activities (i.e., categories) and POIs plays an essential role in next POI prediction; meanwhile, users' activities showcase a noticeable temporal pattern~\cite{zhang2020interactive}. The category encoder thus seeks to capture the activity transition regularity by fusing the temporal context. 
Consequently, the enhanced embeddings of categories in the trajectory $\mathcal{C}^{u,i}$ is, 
\begin{equation}
    \mathbf{C}^{u,i} = \begin{bmatrix}  
    \mathbf{c}^u_{t_1} \mathbf{W}_c+\mathbf{t}^u_1 \mathbf{W}_{t,c}+\mathbf{p}_1 \\ 
    \mathbf{c}^u_{t_2} \mathbf{W}_c+\mathbf{t}^u_2 \mathbf{W}_{t,c}+\mathbf{p}_2\\ \cdots\\
    \mathbf{c}^u_{t_k} \mathbf{W}_c+\mathbf{t}^u_k \mathbf{W}_{t,c}+\mathbf{p}_k \end{bmatrix},
\end{equation}
where $\mathbf{c}^u_{t_k}\in \mathbf{C}$ is the embedding of $c^u_{t_k}$ in $\mathcal{C}^{u,i}$; and $\mathbf{C} \in \mathbb{R}^{\vert\mathcal{C}\vert\times d}$ is the category embedding matrix. Similar as the region encoder, we obtain $\mathbf{F}^{u,i}_c$ as the refined embeddings of categories in $\mathcal{C}^{u,i}$.

\subsection{Region-aware Weighting Strategy} 
As emphasized, user preference on POIs can be highly affected by their preference on regions (\textit{see observations in Section~\ref{sec:data_and_analysis}}) and  categories~\cite{zhang2020interactive, sun2021point}. We, therefore, propose a region-aware weighting strategy to aggregate the signals learned from the three channels for more accurate next POI prediction. However, we notice that there may be contradictory predictions generated by the three channels, that is, the predicted POI, region and category may not always match with each other. To mitigate this issue, it is essential to dynamically adapt the impact of different channels. Intuitively, simpler region check-in patterns (e.g., a trajectory with POIs in a same region only) are 
easier to fit and guarantee more reliable predictions by the region encoder, thus should contribute more than the other encoders on the final predictions, \textit{vice versa}. 

Hence, we split all trajectories with different region check-in patterns into two groups, where one is the trajectories with check-in records in the same region named \textit{same-region group} $G_s$ (e.g., the trajectories of Alice and Bob in Fig.~\ref{fig:running_example}), and the other is the trajectories with cross-region records called \textit{cross-region group} $G_c$ (e.g., the trajectory of Cary in Fig.~\ref{fig:running_example}). 
We believe the region check-in patterns in the same-region group should be easier to fit, thus should be more reliable and contribute more to the final predictions; whereas an opposite situation should be held with the cross-region group. Hence, we adaptively balance the contributions (weights) of different channels for each group. The aggregated representation of user preference on POIs at $t_{k+1}$ is formulated as,  
\begin{equation}
\begin{split}
   \mathbf{f}_{t_{k+1}} = w^{G}_{f_l}\mathbf{f}^l_{t_{k}} + w^{G}_{f_r}\mathbf{f}^r_{t_{k}} +w^{G}_{f_c}\mathbf{f}^c_{t_{k}}, \\
   s.t. w^{G}_{f_l}+w^{G}_{f_r}+w^{G}_{f_c}=1,
\end{split}
\end{equation}
where $w^{G}_{f_l}, w^{G}_{f_r}, w^{G}_{f_c}$ are the learnable weights of location, region and category channels in group $G=\{G_s, G_c\}$; we argue the weights should vary a lot in $G_s$ and $G_c$; and 
$\mathbf{f}^l_{t_{k}}\in\mathbf{F}^{u,i}_l$, $\mathbf{f}^c_{t_{k}}\in\mathbf{F}^{u,i}_c$, $\mathbf{f}^r_{t_{k}}\in\mathbf{F}^{u,i}_r$ are the embeddings of location, category and region at time $t_{k}$.

\subsection{Model Prediction}
Given the learn representation of user preference on POIs at $t_{k+1}$, i.e., $\mathbf{f}_{t_{k+1}}$, the next location can be predicted as follows,
%
\begin{equation}
    \hat{\mathbf{y}}^l_{t_{k+1}} = softmax(\mathbf{f}_{t_{k+1}} \cdot \mathbf{H}^{(Z)}),
\end{equation}
where $\hat{\mathbf{y}}^l_{t_{k+1}}$ represents the prediction probabilities over all candidate POIs; and $Z$ is the total GCN layers. Hence, the loss function for next POI prediction is defined as:
\begin{equation}
\mathcal{J}_l= -\sum\nolimits \mathbf{y}^l_{t_{k+1}}\text{log} (\hat{\mathbf{y}}^l_{t_{k+1}}) + (1-\mathbf{y}^l_{t_{k+1}}) \text{log} (1-\hat{\mathbf{y}}^l_{t_{k+1}}),
\end{equation}
where $\mathbf{y}^l_{t_{k+1}}$ is the one-hot vector of the ground-truth POI $l_{t_{k+1}}$ at $t_{k+1}$.
%
Meanwhile, the two auxiliary region and category channels predict the next region and category, respectively, 
\begin{equation}
\begin{split}
    \hat{\mathbf{y}}^r_{t_{k+1}} = softmax(\mathbf{f}^r_{t_{k}} \cdot \mathbf{R}); 
    \\
    \hat{\mathbf{y}}^c_{t_{k+1}} = softmax(\mathbf{f}^c_{t_{k}} \cdot \mathbf{C}),     
\end{split}
\end{equation}
where $\hat{\mathbf{y}}^r_{t_{k+1}}$ and $\hat{\mathbf{y}}^c_{t_{k+1}}$ are the prediction probabilities over all candidate regions and categories, respectively. The loss functions of the two channels are defined as:
\begin{equation}
\begin{split}
    \mathcal{J}_r= -\sum\nolimits \mathbf{y}^r_{t_{k+1}}\text{log} (\hat{\mathbf{y}}^r_{t_{k+1}}) + (1-\mathbf{y}^r_{t_{k+1}}) \text{log} (1-\hat{\mathbf{y}}^r_{t_{k+1}}),\\
        \mathcal{J}_c= -\sum\nolimits \mathbf{y}^c_{t_{k+1}}\text{log} (\hat{\mathbf{y}}^c_{t_{k+1}}) + (1-\mathbf{y}^c_{t_{k+1}}) \text{log} (1-\hat{\mathbf{y}}^c_{t_{k+1}}),
\end{split}
\end{equation}
where $\mathbf{y}^r_{t_{k+1}}$ and $\mathbf{y}^c_{t_{k+1}}$ are one-hot vectors of the ground truth region $r_{t_{k+1}}$ and category $c_{t_{k+1}}$ at $t_{k+1}$.
Ultimately, we train MCMG by minimizing the following loss function:
\begin{equation}
\mathcal{J}= \mathcal{J}_l + \mathcal{J}_r + \mathcal{J}_c.
\end{equation}

%% file: table/notation.tex
\begin{table}
\renewcommand\arraystretch{1.1}
\centering
\caption{Important notations and descriptions.}
\vspace{-0.1in}
\addtolength{\tabcolsep}{-4pt}
\begin{tabular}{l|l}
\specialrule{.15em}{.15em}{.15em}
\textbf{Notations}  & \textbf{Descriptions}\\ 
\specialrule{.1em}{.1em}{.1em}
$u, l, r, c, g, g' $ & user, POI, region, category, geo-location of POI and region\\
$\mathcal{U}$, $\mathcal{L}$, $\mathcal{R}$, $\mathcal{C}$ & sets of users, POIs, regions, and categories\\
$L^{u}_{t_k}=(l^{u}_{t_k},t^{u}_{k},g^{u}_{t_k})$ & the check-in record of POI performed by user $u$ at time $t_k$\\
$R^{u}_{t_k}=(r^{u}_{t_k},t^{u}_{k},g'^{u}_{t_k})$ & the check-in record of region performed by user $u$ at time $t_k$\\
$C^{u}_{t_k}=(c^{u}_{t_k},t^{u}_{k})$ & the check-in record of category performed by user $u$ at time $t_k$\\
$\mathcal{L}^{u,i}= \{L_{t_1}^u,\cdots, L_{t_k}^u$\} & the $i$-th POI trajectory of user $u$\\
$\mathcal{R}^{u,i}= \{R_{t_1}^u,\cdots, R_{t_k}^u$\} & the $i$-th region trajectory of user $u$\\
$\mathcal{C}^{u,i}= \{C_{t_1}^u,\cdots, C_{t_k}^u$\} & the $i$-th category trajectory of user $u$\\
$d$ &the embedding size\\
$z, Z$ & the $z$-th layer of GCN; and the total layers of GCN\\
$\mathbf{h},\mathbf{r},\mathbf{c} \in \mathbb{R}^d$  & embeddings of POI, region and category\\
$\mathbf{d}^u_k, \mathbf{t}^u_k, \mathbf{p}_k \in \mathbb{R}^d$ & embeddings of distance $d^u_k$, timestamp $t^u_k$ and position $k$\\
$\mathbf{H}^{(z)}\in \mathbb{R}^{\vert\mathcal{L}\vert\times d}$ & the POI embedding matrix in the $z$-th layer of GCN\\
$\mathbf{R} \in \mathbb{R}^{\vert\mathcal{R}\vert\times d}$, $\mathbf{C} \in \mathbb{R}^{\vert\mathcal{C}\vert\times d}$ & embedding matrices of region and category\\
$\mathbf{H}^{u, i}\in\mathbb{R}^{k\times d}$ & embeddings of all POIs within a trajectory $\mathcal{L}^{u, i}$\\
$\mathbf{S}^{u,i}_l\in\mathbb{R}^{k\times d}$& refined embeddings of POIs in $\mathcal{L}^{u,i}$ via the self-attention network\\
$\mathbf{F}^{u,i}_l\in\mathbb{R}^{k\times d}$ & refined embeddings of POIs in $\mathcal{L}^{u,i}$ via the feed-forward network\\
$G_s,G_c$ & same-region group $G_s$ and cross-region group $G_c$\\
$w^{G}_{f_l}, w^{G}_{f_r}, w^{G}_{f_c}$ & weights of POI, region and category channels within group $G=\{G_s,G_c\}$\\
$\hat{\mathbf{y}}^l_{t_{k+1}}$, $\hat{\mathbf{y}}^r_{t_{k+1}}$, $\hat{\mathbf{y}}^c_{t_{k+1}}$  & prediction probabilities over all candidate POIs, regions, and categories \\
$\mathbf{y}^l_{t_{k+1}}$, $\mathbf{y}^r_{t_{k+1}}$, $\mathbf{y}^c_{t_{k+1}}$ & one-hot vectors of ground-truth POI $l_{t_{k+1}}$, region $r_{t_{k+1}}$, and category $c_{t_{k+1}}$\\
\specialrule{.15em}{.15em}{.15em}
\end{tabular}\label{table:notations}
\end{table}

%% file: section/experiments.tex
\section{Experiments and Analysis}\label{subsec:experimental-setup}

\subsection{Datasets}
We adopt Foursquare check-in records from four cities, namely Calgary (CAL), Phoenix (PHO), Singapore (SIN) and New York (NY) as introduced in Section~\ref{subsec:data-overview}. 
Following~\cite{zhang2020interactive}, for each user, we chronologically divide his check-in records into different trajectories by day, and then take the earlier 80\% of his trajectories as training set; the latest 10\% of trajectories as the test set; and the rest 10\% as the validation set. 
Besides, we filter out users with less than three trajectories on the four datasets. 
Our code and data are available at \url{https://github.com/2022MCMG/MCMG}


\subsection{Compared Baselines}
We compare MCMG with five types of baselines.
The first type is the classical method, including,
\begin{itemize}
    \item \textbf{MostPop} recommends next POI with the highest popularity;
    \item \textbf{BPRMF}~\cite{rendle2012bpr} is the Bayesian personalized ranking model based on matrix factorization.
\end{itemize}
The second type is the RNN-based method, including, 
\begin{itemize}
    \item \textbf{ST-RNN}~\cite{liu2016predicting} fuses spatial and temporal contexts into RNN for next POI prediction; 
    \item \textbf{ATST-LSTM}~\cite{huang2019attention} integrates spatial and temporal contexts in a multi-modal manner and applies the attention mechanism for next POI prediction.
\end{itemize}
The third type is the multi-task learning based method, including,
\begin{itemize}
    \item \textbf{MCARNN}~\cite{liao2018predicting} recommends next activity and POI via fusing spatial, time interval and POI category information;
    \item \textbf{iMTL}~\cite{zhang2020interactive} explores the interplay between activity and location for next POI prediction via an interactive manner, which considers spatial, temporal and POI category information.
\end{itemize}
The fourth type is the self-attention based method, namely,
\begin{itemize}
    \item \textbf{SASRec}~\cite{kang2018self} uses the self-attention mechanism for  sequential item recommendation~\cite{fang2020deep}.
\end{itemize}
The last type is the GNN based method, including, 
\begin{itemize}
    \item \textbf{LightGCN}~\cite{he2020lightgcn} is a light version of GCN without feature transformation and nonlinear activation for general item recommendation;
    \item \textbf{SGRec}~\cite{li2021discovering} adopts GAT to extract patterns from the global all-user check-in trajectories facilitated by POI category information, and then uses the attention network to learn user preference from the local individual-user check-in trajectories for next POI prediction.
\end{itemize}

\input{table/hyper_parameter}
\input{table/comparison_results}

\input{table/group_attention}

\subsection{Evaluation Metrics.}
Following state-of-the-arts~\cite{li2021discovering, wang2020next}, hit rate (HR) and normalized discounted cumulative gain (NDCG) are adopted to evaluate the model performance. Generally, higher metric values indicate better ranking accuracy. For a robust comparison, we run each method 5 times, and report the averaged results in Table~\ref{tab:comparison_results}. 

\subsection{Hyper-parameter Settings}
For all the methods, we adopt Bayesian HyperOpt (30 trails) to perform hyper-parameter optimization w.r.t. NDCG@10 on the validation set~\cite{sun2020we}. For our MCMG, the learning rate and regularization coefficient are searched in the range of [0.0001, 0.01]; the embedding size is searched from \{60, 90, 120, 150, 180, 210\}; the batch size is searched from \{256, 512, 1024\}; the heads and blocks of self-attention, as well as the layers of GCN are searched from \{1, 2, 3\}; and the dropout rate for the self-attention module and GCN are searched from $\{0.25, 0.5, 0.75\}$. The detailed parameter searching space and best parameter settings for all methods are reported in Table~\ref{tab:appendix}.

\subsection{Results and Analysis}

\noindent\textbf{Comparative Results and Analysis.}
The performance of all comparison methods regarding HR@$N$ and NDCG@$N$ ($N=\{5, 10\}$) is presented in Table~\ref{tab:comparison_results}, where several major findings can be noted. 
(1) The classical non-sequential methods (MostPop and BPRMF) achieve the most unfavorable performance across the four datasets, which implies the importance of capturing sequential patterns for more accurate next POI recommendation. (2) The multi-task learning based methods (MCARNN and iMTL) outperform RNN based methods (ST-RNN and ATST-LSTM), as they infer user preference on both activities (categories) and POIs. (3) SASRec generally exceeds all the methods mentioned above, showing the superiority of the self-attention mechanism to model sequential patterns. (4) GNN-based methods (i.e., LightGCN and SGRec) consider the global POI check-in trajectories. LightGCN, though being a non-sequential model, can sometimes defeat sequential models (e.g., MCARNN on PHO). It suggests the efficacy of GNN on modelling the global level check-ins for better next POI prediction. By further fusing local POI check-in trajectories and POI categories, the sequential model SGRec far exceeds the other baselines. (5) Our proposed MCMG consistently gains the best performance among all methods. The average improvement over the runner up in terms of HR and NDCG on the four datasets are 9.7\% and 14.2\%, respectively, which firmly confirms the design efficacy of our proposed MCMG. 
\medskip\noindent\textbf{Detailed Analysis on Region-aware Weighting Strategy.}
Table~\ref{tab:group_attention} shows the results of the three channels in different groups regarding region check-in patterns.
The performance of the `entire data' is the average result of the same- and cross-region groups. 
For the ranking accuracy, we find that (a) the region channel in all groups gains the best results in comparison with category and location channels, indicating the more predictable user behavior patterns on the regions of visited POIs; and (b) all channels in the same-region group significantly outperform the corresponding channels in the cross-region group, which verifies that the simpler region check-in pattern is much easier to fit and guarantees a more accurate next region prediction, thus better facilitating next POI recommendation. 
W.r.t. the contributions (i.e., weights) of different channels, several observations can be noted. (a) Generally, the contribution of location channel is smaller than region and category channels; meanwhile, there is only minor variance on the contribution of the location channel between same- and cross-region groups compared with the other two channels. This indicates the importance of region and category on next POI recommendation. 
(b) In the same-region group, the region channel contributes twice as large as the category channel ; whereas the opposite situation is held in the cross-region group. This is mainly attributed to that simpler region check-in pattern guarantees more accurate predictions by the region channel, thus impacting more than the category channel on the final prediction, \textit{vice versa}. 
(c) The average contribution of region channel is larger than that of category channel on the `entire data'. This is because around 80\% of the `entire data' comes from the same-region group as indicated by \textbf{Q1} in Section~\ref{subsec:explore-behavior}. 

\begin{figure}[t]
\centering
\includegraphics[width=0.8\linewidth]{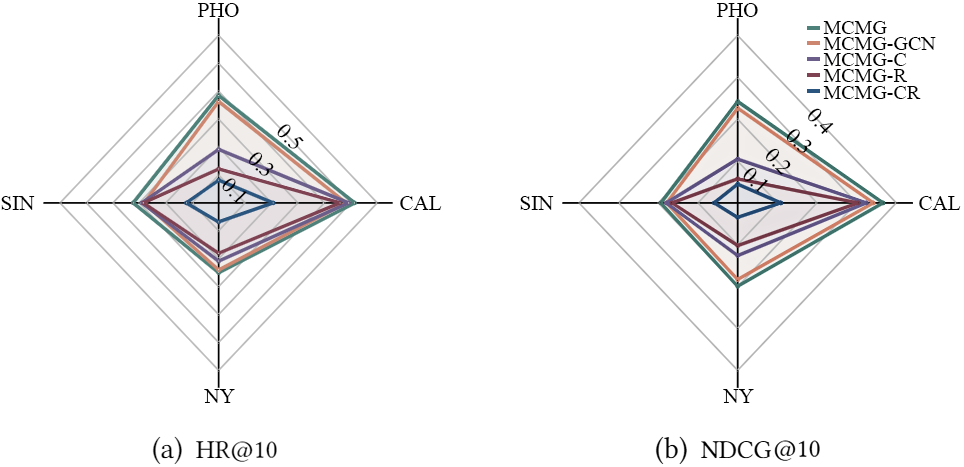}
\caption{Results for different variants of MCMG.}\label{fig:component_ablation}
\end{figure}

\begin{figure}[t]
\centering
\includegraphics[width=0.8\linewidth]{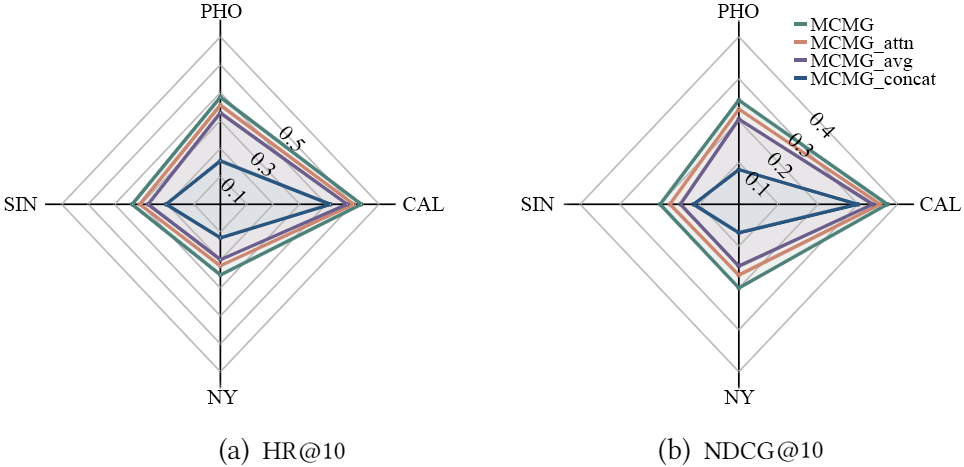}
\caption{Results for different aggregation strategies.}\label{fig:weighting_strategy}
\end{figure}
\medskip\noindent\textbf{Impacts of Different Components.}
We now examine the efficacy of different components of MCMG via comparing with its four variants. (1) MCMG-GCN removes the GCN-based global behavior encoder; (2) MCMG-C removes the auxiliary category channel; (3) MCMG-R removes the auxiliary region channel; and (4) MCMG-CR removes both auxiliary channels. The results are present in Fig.~\ref{fig:component_ablation}, where we can find that (1) MCMG surpasses all the variants, implying the efficacy of each component; (2) MCMG-GCN outperforms all other variants and the performance degradation of MCMG-CR is larger than that of both MCMG-C and MCMG-R, highlighting the great importance of both region and category; and (3) MCMG-C defeats MCMG-R, which emphasizes region plays a more essential role than category for more accurate next POI recommendation. 

\input{plot/parameter_sensitivity}

To verify the efficacy of our region-aware weighting strategy, we compare it with three aggregation strategies, that does not distinguish the same- and cross-region groups.
(1) MCMG\_attn directly adopts an attention network to aggregate the representations learned from the three local channels; (2) MCMG\_avg aggregates the representations learned from the three local channels via average pooling; and (3) MCMG\_concat simply concatenates the representations learned from the three local channels. 
The results are reported in Fig.~\ref{fig:weighting_strategy}, where several findings can be observed. First, MCMG\_concat performs the worst proving that weighted aggregation 
is more effective for next location prediction. Second, both MCMG and MCMG\_attn exceed MCMG\_avg, suggesting the superiority of attention networks. Lastly, MCMG outperforms all other strategies across the four datasets, which reveals the necessity and importance of distinguishing different region check-in patterns on the channel aggregation. 

\medskip\noindent\textbf{Impacts of Important Hyper-parameters.}
Fig.~\ref{fig:parameter_sensitivity} depicts the impact of essential hyper-parameters on MCMG, including the heads and blocks of self-attention, the layers of GCN and the size of embeddings. From the figure, we can observe that (1) the performance of MCMG gradually decreases as the heads of self-attention increases; (2) similar trends can be observed when stacking more blocks of self-attention and GCN layers, and this is because too many blocks and layers may cause the over-fitting issue, thus hurting the performance of MCMG; and (3) the performance first climbs up to reach its peak and then slightly drops with further increase of the size of embeddings, as oversized embeddings may over-represent the
POIs, thus introducing noise.


%% file: table/hyper_parameter.tex
\begin{table*}[t!]
\renewcommand\arraystretch{0.75}
\centering
\footnotesize
\addtolength{\tabcolsep}{-2pt}
\caption{The search space of hyper-parameters and the optimal settings found by Bayesian HyperOpt for all methods on the four real-world datasets.}\label{tab:appendix}
\vspace{-0.1in}
\begin{tabular}{c|l|l|l|l|l|l|l}
\specialrule{.15em}{.05em}{.05em}
\multicolumn{1}{l|}{}  & Parameter        & CAL    & PHO    & SIN    & NY     & Searching space                & Description                                       \\ \specialrule{.1em}{.1em}{.1em}
\multirow{5}{*}{\rotatebox{90}{BPRMF}}  & -embedding\_size & 60     & 60     & 60     & 60     & \{60, 90, 120, 150, 180, 210\} & the size of embeddings                   \\
 
  & -lr  & 0.0093   & 0.0084   & 0.0091   & 0.0089   & {[}0.0001, 0.01{]}   & learning rate  \\
& -lambda   & 0.0050   & 0.0081   & 0.0093   & 0.0091   & {[}0.0001, 0.01{]}             & L2 regularization coefficient \\
& -batch\_size  & 256    & 256    & 512    & 512    & \{256, 512, 1024\}             & batch size\\ \specialrule{.05em}{.05em}{.05em}
\multirow{6}{*}{\rotatebox{90}{ST-RNN}}  & -embedding\_size & 90     & 90     & 120    & 120    & \{60, 90, 120, 150, 180, 210\} & the size of embeddings                   \\
& -lr  & 0.0091   & 0.0086   & 0.0097   & 0.0092   & {[}0.0001, 0.01{]}   & learning rate                                     \\
& -lambda   & 0.0015  & 0.0026  & 0.0042  & 0.0013  & {[}0.0001, 0.01{]}& L2 regularization coefficient                     \\
 & -batch\_size & 256    & 512    & 512    & 512    & \{256, 512, 1024\} & batch size                                        \\
& -layers & 1   & 1    & 1      & 1      & \{1, 2, 3\}   & the number of RNN layers                           \\ \specialrule{.05em}{.05em}{.05em}
\multicolumn{1}{l|}{\multirow{6}{*}{\rotatebox{90}{ATSTLSTM}}} & -embedding\_size & 60     & 120    & 120    & 120    & \{60, 90, 120, 150, 180, 210\} & the size of embeddings                   \\
\multicolumn{1}{l|}{}    & -lr              & 0.0009 & 0.0010 & 0.0034 & 0.0012 & {[}0.0001, 0.01{]}             & learning rate                                     \\
\multicolumn{1}{l|}{}   & -lambda          & 0.0004  & 0.0004  & 0.0003  & 0.0002  & {[}0.0001, 0.01{]}             & L2 regularization coefficient                     \\
\multicolumn{1}{l|}{}      & -batch\_size     & 512    & 512    & 512    & 1024   & \{256, 512, 1024\}             & batch size                                        \\
\multicolumn{1}{l|}{}   & -layers  & 1   & 1      & 1      & 1      & \{1, 2, 3\}  & the number of LSTM layers                          \\ \specialrule{.05em}{.05em}{.05em}
\multirow{6}{*}{\rotatebox{90}{MCARNN}}                          & -embedding\_size & 90     & 90     & 120    & 120    & \{60, 90, 120, 150, 180, 210\} & the size of embeddings                   \\
& -lr  & 0.0089   & 0.0091   & 0.0085   & 0.0085   & {[}0.0001, 0.01{]}             & learning rate                                     \\
  & -lambda  & 0.0091   & 0.0079   & 0.0083   & 0.0094   & {[}0.0001, 0.01{]} & L2 regularization coefficient                     \\
 & -batch\_size   & 512  & 512  & 512 & 512    & \{256, 512, 1024\}             & batch size                                        \\
 & -layers   & 2  & 2  & 2  & 2 & \{1, 2, 3\}   & the number of RNN layers \\ \specialrule{.05em}{.05em}{.05em}
\multirow{6}{*}{\rotatebox{90}{iMTL}}  & -embedding\_size & 120    & 120    & 120    & 120    & \{60, 90, 120, 150, 180, 210\} & the size of embeddings\\
& -lr  & 0.0001 & 0.0001 & 0.0001 & 0.0002 & {[}0.0001, 0.01{]}  & learning rate                                     \\
 & -lambda   & 0.0001 & 0.0001 & 0.0001 & 0.0001 & {[}0.0001, 0.01{]}             & L2 regularization coefficient                     \\
 & -batch\_size  & 256  & 512 & 512  & 512 & \{256, 512, 1024\}             & batch size                                        \\
& -layers & 2  & 2 & 2  & 2  & \{1, 2, 3\}  & the number of LSTM layers          \\ 
\specialrule{.05em}{.05em}{.05em}
\multirow{8}{*}{\rotatebox{90}{SASRec}}                          & -embedding\_size & 60     & 90     & 90     & 120    & \{60, 90, 120, 150, 180, 210\} & the size of embeddings   \\
& -lr  & 0.0005 & 0.0008 & 0.0013 & 0.0016  & {[}0.0001, 0.01{]}             & learning rate                                     \\
& -lambda   & 0.0001 & 0.0002 & 0.0002 & 0.0002 & {[}0.0001, 0.01{]}     & L2 regularization coefficient                     \\
& -batch\_size & 256& 512 & 512 & 512  & \{256, 512, 1024\}  & batch size \\
& -heads   & 1   & 1 & 1   & 1 & \{1, 2, 3\}            & the heads of self-attention\\
& -blocks   & 1 & 1& 1  & 1 & \{1, 2, 3\}            & the blocks of self-attention                      \\
 & -dropout      & 0.25   & 0.25   & 0.25   & 0.75   & \{0.25, 0.5, 0.75\}  
 & the dropout rate of self-attention                \\ \specialrule{.05em}{.05em}{.05em}
\multirow{7}{*}{\rotatebox{90}{LightGCN}}  & -embedding\_size & 90     & 90     & 90     & 120    & \{60, 90, 120, 150, 180, 210\} & the size of embeddings \\   
& -lr & 0.0018  & 0.0014  & 0.0012  & 0.0011  & {[}0.0001, 0.01{]}  & learning rate \\
& -lambda  & 0.0001 & 0.0004 & 0.0002 & 0.0002 & {[}0.0001, 0.01{]}    & L2 regularization coefficient   \\
& -batch\_size & 512    & 512    & 512    & 512    & \{256, 512, 1024\}     & batch size \\
& -dropout         & 0.25   & 0.25   & 0.5    & 0.5    & \{0.25, 0.5, 0.75\}   & the dropout rate of GCN \\  & -layers  & 2 & 2 & 2 & 2  & \{1, 2, 3\}  & the number of GCN layers                       \\ \specialrule{.05em}{.05em}{.05em}
\multirow{7}{*}{\rotatebox{90}{SGRec}}                           & -embedding\_size & 120    & 120    & 120    & 120    & \{60, 90, 120, 150, 180, 210\} & the size of embeddings                   \\
 & -lr              & 0.0051  & 0.0045  & 0.0091   & 0.0099   & {[}0.0001, 0.01{]}              & learning rate                                     \\
 & -lambda          & 0.0001 & 0.0002 & 0.0002 & 0.0004 & {[}0.0001, 0.01{]}              & L2 regularization coefficient                     \\
 & -batch\_size   & 512  & 512  & 512    & 1024   & \{256, 512, 1024\}             & batch size                                        \\
 & -dropout & 0.25   & 0.25  & 0.25   & 0.25   & \{0.25, 0.5, 0.75\}            & the dropout rate of GAT                                                  \\
 & -layers  & 1   & 1  & 1   & 1 & \{1, 2, 3\}                    & the number of GAT layers                                                 \\ \specialrule{.05em}{.05em}{.05em}
\multirow{10}{*}{\rotatebox{90}{MCMG}}                           & -embedding\_size & 180    & 120    & 120    & 120    & \{60, 90, 120, 150, 180, 210\} & the size of embeddings                   \\
& -lr              & 0.0043  & 0.0024  & 0.0016  & 0.0015  & {[}0.0001, 0.01{]}              & learning rate                                     \\
& -lambda          & 0.0001 & 0.0001 & 0.0001 & 0.0001 & {[}0.0001, 0.01{]}              & L2 regularization coefficient                     \\
& -batch\_size     & 512    & 512    & 512    & 1024   & \{256, 512, 1024\}             & batch size                                        \\
& -layers       & 1      & 1      & 1      & 1      & \{1, 2, 3\}                    & the number of GCN layers                       \\
& -heads         & 1      & 1      & 1      & 1      & \{1, 2, 3\}                    & the heads of self-attention\\
& -blocks        & 1      & 1      & 1      & 1      & \{1, 2, 3\}                    & the blocks of self-attention                      \\
& -GCN\_dropout    & 0.5    & 0.5    & 0.5    & 0.25   & \{0.25, 0.5, 0.75\}            & the dropout rate of GCN  \\
& -SA\_dropout     & 0.5    & 0.5    & 0.5    & 0.75   & \{0.25, 0.5, 0.75\}  & the dropout rate of self-attention  \\
\specialrule{.15em}{.05em}{.05em}
\end{tabular}
\end{table*}

%% file: table/comparison_results.tex
\begin{table*}[t]
\renewcommand\arraystretch{1.1}
\centering
\addtolength{\tabcolsep}{-3.8pt}
\caption{Performance of all methods on the four real-world datasets. The best performance is boldfaced; the runner up is underlined; and `Improve' shows improvements achieved by MCMG relative to the runner up, whose significance is determined by a paired t-test (** for $p<0.01$ and *** for $p<0.001$).
}\label{tab:comparison_results}
\vspace{-0.1in}
	\begin{tabular}{l|c|c|c|c|c|c|c|c} 
	\specialrule{.15em}{.05em}{.05em}
		& \multicolumn{4}{c|}{CAL}
		& \multicolumn{4}{c}{PHO}
		\\ \cline{2-9}
		& HR@5  & HR@10  & NDCG@5 & NDCG@10 & HR@5  & HR@10  & NDCG@5 & NDCG@10  \\ 
		\specialrule{.05em}{.05em}{.05em}
		MostPop & 0.0622  &  0.0913 & 0.0375  & 0.0463  & 0.0160  &  0.0223  &  0.0114  & 0.0131    \\
		BPRMF & 0.0862  & 0.1046  & 0.0467  & 0.0793  &  0.0487 &  0.0585  &  0.0256  &  0.0304   \\
		ST-RNN & 0.1469  & 0.1731  & 0.1225  & 0.1508  &  0.1240 &  0.2028  &  0.0802  &  0.1229   \\
		ATST-LSTM &  0.2027 &  0.2898 &  0.1684 & 0.2236   & 0.1579  &  0.2377  & 0.1033   &  0.1385   \\
		MCARNN & 0.2451  & 0.3286  & 0.2015  &  0.2693 & 0.1905  & 0.2726   &  0.1264  &  0.1617   \\
		iMTL & 0.2216  & 0.3104  & 0.2031  & 0.2545  &  0.1830 &  0.2747  &  0.1301  &  0.1632   \\
		SASRec & 0.3077  & 0.4108  & 0.2646   & 0.2723  & 0.2807  &  0.3325  & 0.2021   &  0.2101   \\
		LightGCN & 0.2954  &  0.3731 &  0.1868 &  0.2076 & 0.2563  &  0.3151  & 0.1881   &  0.2194   \\
		SGRec &  \underline{0.3879} & \underline{0.4854}  &  \underline{0.2894} & \underline{0.3112}  &  \underline{0.2897} &  \underline{0.3401}  &  \underline{0.2048}  &  \underline{0.2249}   \\
		\specialrule{.05em}{.05em}{.05em}
		MCMG & \textbf{0.4426 }$^{**}$  & \textbf{0.5333 }$^{**}$  & \textbf{0.3431 }$^{**}$  &  \textbf{0.3743 }$^{**}$ & \textbf{ 0.3027}$^{**}$  &  \textbf{0.3843 }$^{**}$  &  \textbf{ 0.2211}$^{**}$  &   \textbf{0.2489 }$^{**}$  \\
		Improve &  14.1\% &  9.9\% &  18.6\% &  20.3\% &  4.5\% &  13.0\%  &  8.0\%  &  10.7\%   \\
		\specialrule{.15em}{.05em}{.05em}
		\specialrule{.15em}{.05em}{.05em}
		& \multicolumn{4}{c|}{SIN}
		& \multicolumn{4}{c}{NY}
		\\ \cline{2-9}
		& HR@5  & HR@10  & NDCG@5 & NDCG@10 & HR@5  & HR@10  & NDCG@5 & NDCG@10  \\ 
		\specialrule{.05em}{.05em}{.05em}
		MostPop &  0.0125 &  0.0293 & 0.0106   & 0.0151  & 0.0139  &  0.0235  &  0.0101  &  0.0145   \\
		BPRMF &  0.0352 &  0.0525 & 0.0222  & 0.0380  &  0.0381 &  0.0468  &  0.0239  &  0.0339    \\
		ST-RNN & 0.0959  & 0.1370  & 0.0655  & 0.0794   &  0.1347 &  0.1826   &  0.0593  &  0.1303   \\
		ATST-LSTM & 0.1296  & 0.1933 & 0.1027  & 0.1476  &  0.1667 &  0.2031  &  0.0912  &  0.1638    \\
		MCARNN &  0.2018 & 0.2692  &  0.1169  & 0.1591  & 0.1835  &  0.2397  &  0.1036  &  0.1870   \\
		iMTL &  0.1505 & 0.1801  &  0.1051 &  0.1423 & 0.1798  &  0.2422  &  0.0989  &  0.1861   \\
		SASRec & 0.2301  & 0.2885  & 0.1301  & 0.1524  &  0.1815  & 0.2401   &  0.1021  &  0.1398   \\
		LightGCN & 0.2165  &  0.2691 & 0.1263   &  0.1335 & 0.1752  &  0.2229  &  0.1035  &  0.1209   \\
		SGRec & \underline{0.2310 } & \underline{0.2953 }  &  \underline{0.1530 } & \underline{0.1739 }  &  \underline{0.1891 } &  \underline{0.2443 }  &  \underline{0.1089 }  &  \underline{0.1877 }     \\
		\specialrule{.05em}{.05em}{.05em}
		MCMG & \textbf{0.2498 }$^{**}$  & \textbf{0.3338 }$^{**}$  & \textbf{0.1729 }$^{***}$  &  \textbf{0.1987 }$^{***}$ & \textbf{0.2102 }$^{**}$  &  \textbf{0.2531 }$^{**}$  &  \textbf{0.1329 }$^{**}$  &   \textbf{0.1997 }$^{***}$  \\
		Improve &  8.1\% &  13.0\% &  13.0\% &  14.3\% &  11.2\% &  3.6\%  &  22.4\%  &  6.4\%   \\
        \specialrule{.15em}{.05em}{.05em}
	\end{tabular}
	\label{table:variation}
	\vspace{-0.1in}
\end{table*}

%% file: table/group_attention.tex
\begin{table*}[!ht]
\renewcommand\arraystretch{1.1}
\centering
\addtolength{\tabcolsep}{-2.5pt}
\caption{The performance of different groups regarding region check-in patterns (i.e., entire data, same-region and cross-region groups) on the three channels (i.e., POI channel, category channel and region channel).}\label{tab:group_attention}
\vspace{-0.1in}
\begin{tabular}{cl|ccc|ccc|ccc}
\specialrule{.15em}{.15em}{.15em}
\multicolumn{2}{c|}{\multirow{2}{*}{}}  & \multicolumn{3}{c|}{entire data}& \multicolumn{3}{c|}{same-region group} & \multicolumn{3}{c}{cross-region group} \\ \cline{3-11} 
\multicolumn{2}{c|}{}  & POI  & category  & region   & POI  & category  & region  & POI  & category  & region   \\
\specialrule{.1em}{.1em}{.1em}
\multicolumn{1}{c|}{\multirow{5}{*}{CAL}} & HR@5 & 0.4426 & 0.4592 & 0.9421 & {0.5033} & {0.5099} & {1.0000} & {0.3720} & {0.4085} & {0.8841}  \\ 
\multicolumn{1}{c|}{} & HR@10 & 0.5333 & 0.5946 & 1.0000 & {0.6093} & {0.6159} & 1.0000 & {0.4573}  & {0.5732} & 1.0000  \\ 
\multicolumn{1}{c|}{}  & NDCG@5 & 0.3431 & 0.3644 & 0.7942 & {0.3903} & {0.4194} & {0.9438} & {0.2906} & {0.3095} & {0.6445} \\ 
\multicolumn{1}{c|}{}  & NDCG@10 & 0.3743 & 0.3994 & 0.8264 & {0.4246} & {0.4349} & {0.9683} & {0.3241} & {0.3639} & {0.6845} \\ 
\multicolumn{1}{c|}{}  & weight & 0.23 & 0.35 & 0.42 & {0.20} & {0.22} & \textbf{0.58} & {0.25} & {0.48} & {0.27}\\ \specialrule{.05em}{.05em}{.05em}
\multicolumn{1}{c|}{\multirow{5}{*}{PHO}} & HR@5 & 0.3027 & 0.4191 & 0.9498 & {0.3805} & {0.4566} & {1.0000} & {0.2250} & {0.3815} & {0.8996}  \\ 
\multicolumn{1}{c|}{} & HR@10 & 0.3843 & 0.5555 & 1.0000 & {0.4893} & {0.5998} & 1.0000 & {0.2793} & {0.5112} & 1.0000  \\ 
\multicolumn{1}{c|}{}  & NDCG@5 & 0.2211 & 0.3015 & 0.8083 &  {0.2791} & {0.3530} & {0.9958} & {0.1631} & {0.2499} & {0.6208} \\ 
\multicolumn{1}{c|}{}  & NDCG@10 & 0.2489 & 0.3591 & 0.8330 & {0.3098} & {0.4183} & {0.9976} & {0.1879} & {0.2999} & {0.6684} \\ 
\multicolumn{1}{c|}{}  & weight & 0.25 & 0.33 & 0.42 & {0.23} & {0.22} & \textbf{0.55} & {0.28} & {0.43} & {0.29}\\ \specialrule{.05em}{.05em}{.05em}
\multicolumn{1}{c|}{\multirow{5}{*}{SIN}} & HR@5 & 0.2498 & 0.4194 & 0.9310 & {0.3210} & {0.4238} & {1.0000} & {0.1787} & {0.4149} & {0.8619}  \\ 
\multicolumn{1}{c|}{} & HR@10 & 0.3338 & 0.5710 & 1.0000 & {0.4198} & {0.5803} & 1.0000 & {0.2479} & {0.5617} & 1.0000 \\ 
\multicolumn{1}{c|}{}  & NDCG@5 & 0.1729  & 0.3097 & 0.7682 & {0.2488} & {0.3607} & {0.9991} & {0.0970} & {0.2586} & {0.5372} \\ 
\multicolumn{1}{c|}{}  & NDCG@10 &  0.1987 & 0.3215 & 0.8136 & {0.2832} & {0.3409} & {0.9995} & {0.1143} & {0.3021} & {0.6277} \\ 
\multicolumn{1}{c|}{}  & weight & 0.22 & 0.37 & 0.41 & {0.21} & {0.25} & \textbf{0.54} & {0.24} & {0.49} & {0.27}\\ \specialrule{.05em}{.05em}{.05em}
\multicolumn{1}{c|}{\multirow{5}{*}{NY}} & HR@5 & 0.2102 & 0.2993 & 0.9514 & {0.2500} & {0.3131} & {1.0000} & {0.1704} & {0.2855} & {0.9027}  \\ 
\multicolumn{1}{c|}{} & HR@10 & 0.2531 & 0.4122 & 1.0000 & {0.2962} & {0.4250} & 1.0000 & {0.2100} & {0.3994}  & 1.0000  \\ 
\multicolumn{1}{c|}{}  & NDCG@5 & 0.1329 & 0.2143 & 0.7939 & {0.1612} & {0.2398} & {0.9993} & {0.1046} & {0.1887} & {0.5884} \\ 
\multicolumn{1}{c|}{}  & NDCG@10 & 0.1997 & 0.2509 & 0.8095 & {0.2399} & {0.2783} & {0.9996} & {0.1595}  & {0.2234} & {0.6193} \\ 
\multicolumn{1}{c|}{}  & weight & 0.27 & 0.35 & 0.38 & {0.23} & {0.21} & \textbf{0.56} & {0.31} & {0.50} & {0.19} \\ 
\specialrule{.15em}{.15em}{.15em}
\end{tabular}
\vspace{-0.1in}
\end{table*}

%% file: plot/parameter_sensitivity.tex
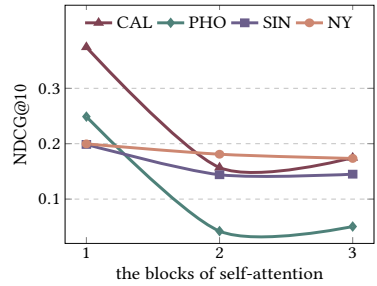
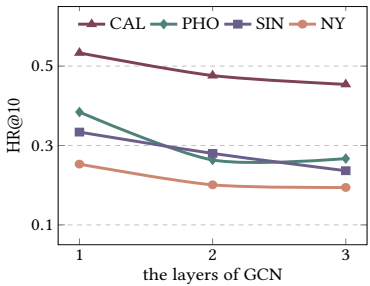
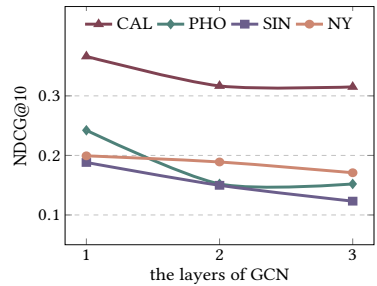
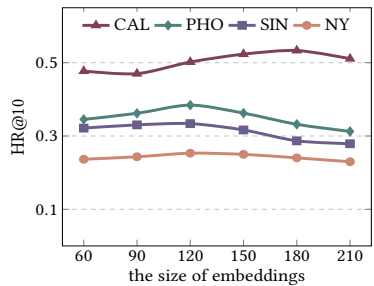
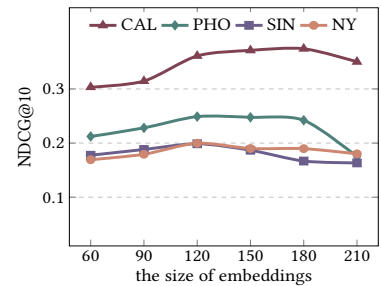
\begin{figure*}[t]
	\centering
	\footnotesize
\subfigure[The impact of heads of self-attention]{
		\begin{tikzpicture}[xscale=0.60,yscale=0.55]
		\begin{axis}[
 		xlabel ={the heads of self-attention},
	    ylabel={HR@10},
		xmin=1, xmax=3,
		ymin=0.05, ymax=0.65,
		xtick={1,2,3},
		yticklabel style={/pgf/number format/.cd,fixed,precision=3},
		ytick={0.1,0.3,0.5},
		ylabel style ={font = \Large},
		xlabel style ={font = \Large}, enlarge x limits= 0.08,
		scaled ticks=false,
		tick label style={font = \Large},
		legend style={at={(0.5,0.98)},font = \Large, anchor=north,legend columns=4,draw=none},
		ymajorgrids=true,
		grid style=dashed,
		]
		\addplot[color=4,
		mark=triangle,
		line width=2pt,
		smooth]coordinates {
			( 1 , 0.5333 )
            ( 2 , 0.4952 )
            ( 3 , 0.4889 )
		};
		\addplot[color=9,
		mark=diamond,
		line width=2pt,
		smooth] coordinates {
    		( 1 , 0.3843 )
            ( 2 , 0.3297 )
            ( 3 , 0.2592 )
		};
			\addplot[color=6,
		mark=square*,
		line width=2pt,
		smooth] coordinates {
			( 1 , 0.3338 )
            ( 2 , 0.3131 )
            ( 3 , 0.3098)
			};	
		\addplot[color=7,
		mark=*,
		line width=2pt,
		smooth] coordinates {
			( 1 , 0.2531)
            ( 2 , 0.2204)
            ( 3 , 0.2019)
		};
			\legend{CAL, PHO, SIN, NY}
		\end{axis}
		\end{tikzpicture}
	}\hspace{0.5in}
\subfigure[The impact of heads of self-attention]{
		\begin{tikzpicture}[xscale=0.60,yscale=0.55]
		\begin{axis}[
		xlabel ={the heads of self-attention},
	    ylabel={NDCG@10},
		xmin=1, xmax=3,
		ymin=0.05, ymax=0.45,
		xtick={1,2,3},
		yticklabel style={/pgf/number format/.cd,fixed,precision=3},
		ytick={0.1,0.2,0.3},
		ylabel style ={font = \Large},
		xlabel style ={font = \Large}, enlarge x limits= 0.08,
		scaled ticks=false,
		tick label style={font = \Large},
		legend style={at={(0.5,0.98)},font = \Large, anchor=north,legend columns=4,draw=none},
		ymajorgrids=true,
		grid style=dashed,
		]
		\addplot[color=4,
		mark=triangle,
		line width=2pt,
		smooth]coordinates {
			( 1 , 0.3743 )
            ( 2 , 0.3435 )
            ( 3 , 0.3073 )
		};
		\addplot[color=9,
		mark=diamond,
		line width=2pt,
		smooth] coordinates {
    		( 1 , 0.2489 )
            ( 2 , 0.2165 )
            ( 3 , 0.2097 )
		};
			\addplot[color=6,
		mark=square*,
		line width=2pt,
		smooth] coordinates {
			( 1 , 0.1987 )
            ( 2 , 0.1878 )
            ( 3 , 0.1882)
			};	
		\addplot[color=7,
		mark=*,
		line width=2pt,
		smooth] coordinates {
			( 1 , 0.1997)
            ( 2 , 0.1993)
            ( 3 , 0.1990)
		};
			\legend{CAL, PHO, SIN, NY}
		\end{axis}
		\end{tikzpicture}
	}
	
	\subfigure[The impact of blocks of self-attention]{
		\begin{tikzpicture}[xscale=0.60,yscale=0.55]
		\begin{axis}[
		xlabel ={the blocks of self-attention},
	    ylabel={HR@10},
		xmin=1, xmax=3,
		ymin=0.02, ymax=0.65,
		xtick={1,2,3},
		yticklabel style={/pgf/number format/.cd,fixed,precision=3},
		ytick={0.1,0.3,0.5},
		ylabel style ={font = \Large},
		xlabel style ={font = \Large}, enlarge x limits= 0.08,
		scaled ticks=false,
		tick label style={font = \Large},
		legend style={at={(0.5,0.98)},font = \Large, anchor=north,legend columns=4,draw=none},
		ymajorgrids=true,
		grid style=dashed,
		]
		\addplot[color=4,
		mark=triangle,
		line width=2pt,
		smooth]coordinates {
			( 1 , 0.5333 )
            ( 2 , 0.3001 )
            ( 3 , 0.2984 )
		};
		\addplot[color=9,
		mark=diamond,
		line width=2pt,
		smooth] coordinates {
    		( 1 , 0.3843 )
            ( 2 , 0.0792 )
            ( 3 , 0.1024 )
		};
			\addplot[color=6,
		mark=square*,
		line width=2pt,
		smooth] coordinates {
			( 1 , 0.3338 )
            ( 2 , 0.2192 )
            ( 3 , 0.1924)
			};
		\addplot[color=7,
		mark=*,
		line width=2pt,
		smooth] coordinates {
			( 1 , 0.2531)
            ( 2 , 0.2019)
            ( 3 , 0.1801 )
		};
			\legend{CAL, PHO, SIN, NY}
		\end{axis}
		\end{tikzpicture}
	}\hspace{0.5in}
	\subfigure[The impact of blocks of self-attention]{
		\begin{tikzpicture}[xscale=0.60,yscale=0.55]
		\begin{axis}[
 		xlabel ={the blocks of self-attention},
	    ylabel={NDCG@10},
		xmin=1, xmax=3,
		ymin=0.02, ymax=0.45,
		xtick={1,2,3},
		yticklabel style={/pgf/number format/.cd,fixed,precision=3},
		ytick={0.1,0.2,0.3},
		ylabel style ={font = \Large},
		xlabel style ={font = \Large}, enlarge x limits= 0.08,
		scaled ticks=false,
		tick label style={font = \Large},
		legend style={at={(0.5,0.98)},font = \Large, anchor=north,legend columns=4,draw=none},
		ymajorgrids=true,
		grid style=dashed,
		]
		\addplot[color=4,
		mark=triangle,
		line width=2pt,
		smooth]coordinates {
			( 1 , 0.3743 )
            ( 2 , 0.157 )
            ( 3 , 0.1747 )
		};
		\addplot[color=9,
		mark=diamond,
		line width=2pt,
		smooth] coordinates {
    		( 1 , 0.2489 )
            ( 2 , 0.0422 )
            ( 3 , 0.0505 )
		};
			\addplot[color=6,
		mark=square*,
		line width=2pt,
		smooth] coordinates {
			( 1 , 0.1987 )
            ( 2 , 0.144 )
            ( 3 , 0.145)
			};
		\addplot[color=7,
		mark=*,
		line width=2pt,
		smooth] coordinates {
			( 1 , 0.1997)
            ( 2 , 0.1810)
            ( 3 , 0.1734 )
		};
			\legend{CAL, PHO, SIN, NY}
		\end{axis}
		\end{tikzpicture}
	}
	
	\subfigure[The impact of layers of GCN]{
		\begin{tikzpicture}[xscale=0.60,yscale=0.55]
		\begin{axis}[
	    xlabel ={the layers of GCN},
		ylabel={HR@10},
		xmin=1, xmax=3,
		ymin=0.05, ymax=0.65,
		xtick={1,2,3},
		yticklabel style={/pgf/number format/.cd,fixed,precision=3},
		ytick={0.1,0.3,0.5},
		ylabel style ={font = \Large},
		xlabel style ={font = \Large}, enlarge x limits= 0.08,
		scaled ticks=false,
		tick label style={font = \Large},
		legend style={at={(0.5,0.98)},font = \Large, anchor=north,legend columns=4,draw=none},
		ymajorgrids=true,
		grid style=dashed,
		]
		\addplot[color=4,
 		mark=triangle,
		line width=2pt,
		smooth]coordinates {
			( 1 , 0.5333 )
			( 2 ,0.4762 )
			( 3 , 0.4540 )
		};\label{CAL}
		\addplot[color=9,
		mark=diamond,
		line width=2pt,
		smooth] coordinates {
			( 1 ,0.3843 )
			( 2 , 0.2636)
			( 3 ,0.2670 )
			};	\label{PHO}
			\addplot[color=6,
		mark=square*,
		line width=2pt,
		smooth] coordinates {
			( 1 ,0.3338 )
			( 2 , 0.2801 )
			( 3 ,0.2365 )
			};	\label{SIN}
		\addplot[color=7,
		mark=*,
		line width=2pt,
		smooth] coordinates {
    		( 1 ,0.2531)
    		( 2 , 0.2009)
    		( 3 , 0.1943)
		};\label{NY}
		\legend{CAL, PHO, SIN, NY}
		\end{axis}
		\end{tikzpicture}
		}\hspace{0.5in}
	\subfigure[The impact of layers of GCN]{
		\begin{tikzpicture}[xscale=0.60,yscale=0.55]
		\begin{axis}[
	    xlabel ={the layers of GCN},
		ylabel={NDCG@10},
		xmin=1, xmax=3,
		ymin=0.05, ymax=0.45,
		xtick={1,2,3},
		yticklabel style={/pgf/number format/.cd,fixed,precision=3},
		ytick={0,0.1,0.2,0.3},
		ylabel style ={font = \Large},
		xlabel style ={font = \Large}, enlarge x limits= 0.08,
		scaled ticks=false,
		tick label style={font = \Large},
		legend style={at={(0.5,0.98)},font = \Large, anchor=north,legend columns=4,draw=none},
		ymajorgrids=true,
		grid style=dashed,
		]
		\addplot[color=4,
 		mark=triangle,
		line width=2pt,
		smooth]coordinates {
			( 1 , 0.3663 )
			( 2 ,0.3167 )
			( 3 , 0.3151 )
		};\label{CAL}
		\addplot[color=9,
		mark=diamond,
		line width=2pt,
		smooth] coordinates {
			( 1 ,0.2422 )
			( 2 , 0.1522 )
			( 3 ,0.1521 )
			};	\label{PHO}
			\addplot[color=6,
		mark=square*,
		line width=2pt,
		smooth] coordinates {
			( 1 ,0.1882 )
			( 2 , 0.1498 )
			( 3 ,0.1232 )
			};	\label{SIN}
		\addplot[color=7,
		mark=*,
		line width=2pt,
		smooth] coordinates {
    		( 1 ,0.1993)
    		( 2 , 0.1890)
    		( 3 , 0.1709)
		};\label{NY}
		\legend{CAL, PHO, SIN, NY}
		\end{axis}
		\end{tikzpicture}
		}
		
\hspace{0.045in}
	\subfigure[The impact of size of embeddings]{
		\begin{tikzpicture}[xscale=0.60,yscale=0.55]
		\begin{axis}[
		xlabel ={the size of embeddings},
	    ylabel={HR@10},
			xmin=60, xmax=210,
		ymin=0, ymax=0.65,
		xtick={60,90,120,150,180,210},
		yticklabel style={/pgf/number format/.cd,fixed,precision=3},
		ytick={0.1,0.3,0.5},
		ylabel style ={font = \Large},
		xlabel style ={font = \Large}, enlarge x limits= 0.08,
		scaled ticks=false,
		tick label style={font = \Large},
		legend style={at={(0.5,0.98)},font = \Large, anchor=north,legend columns=4,draw=none},
		ymajorgrids=true,
		grid style=dashed,
		]
		\addplot[color=4,
		mark=triangle*,
		line width=2pt,
		smooth]coordinates {
			( 60 , 0.4771 )
			( 90 , 0.4701 )
            ( 120 , 0.5019 )
            ( 150 , 0.5234 )
            ( 180 , 0.5333 )
            ( 210 , 0.5110 )
		};
		\addplot[color=9,
		mark=diamond,
		line width=2pt,
		smooth] coordinates {
    		( 60 , 0.3453 )
    		( 90 , 0.3619 )
            ( 120 , 0.3843 )
            ( 150 , 0.3624 )
            ( 180 , 0.3319 )
            ( 210 , 0.3125)
		};
			\addplot[color=6,
		mark=square*,
		line width=2pt,
		smooth] coordinates {
			( 60 , 0.3215 )
			( 90 , 0.3304 )
            ( 120 , 0.3338 )
            ( 150 , 0.3165 )
            ( 180 , 0.2868 )
            ( 210 , 0.2789 )
			};	
		\addplot[color=7,
		mark=*,
		line width=2pt,
		smooth] coordinates {
			( 60 , 0.2365)
			( 90 , 0.2433)
            ( 120 , 0.2531)
            ( 150 , 0.2499 )
            ( 180 , 0.2403 )
            ( 210 , 0.2298 )
		};
			\legend{CAL, PHO, SIN, NY}
		\end{axis}
		\end{tikzpicture}
	}\hspace{0.5in}
	\subfigure[The impact of size of embeddings]{
		\begin{tikzpicture}[xscale=0.60,yscale=0.55]
		\begin{axis}[
		xlabel ={the size of embeddings},
	    ylabel={NDCG@10},
			xmin=60, xmax=210,
		ymin=0.01, ymax=0.45,
		xtick={60,90,120,150,180,210},
		yticklabel style={/pgf/number format/.cd,fixed,precision=3},
		ytick={0.1,0.2,0.3},
		ylabel style ={font = \Large},
		xlabel style ={font = \Large}, enlarge x limits= 0.08,
		scaled ticks=false,
		tick label style={font = \Large},
		legend style={at={(0.5,0.98)},font = \Large, anchor=north,legend columns=4,draw=none},
		ymajorgrids=true,
		grid style=dashed,
		]
		\addplot[color=4,
		mark=triangle*,
		line width=2pt,
		smooth]coordinates {
			( 60 , 0.3033 )
			( 90 , 0.3145 )
            ( 120 , 0.361 )
            ( 150 , 0.3713 )
            ( 180 , 0.3743 )
            ( 210 , 0.3501 )
		};
		\addplot[color=9,
		mark=diamond,
		line width=2pt,
		smooth] coordinates {
    		( 60 , 0.2122 )
    		( 90 , 0.2281 )
            ( 120 , 0.2489 )
            ( 150 , 0.2476 )
            ( 180 , 0.2422 )
            ( 210 , 0.1751)
		};
			\addplot[color=6,
		mark=square*,
		line width=2pt,
		smooth] coordinates {
			( 60 , 0.1775 )
			( 90 , 0.1882 )
            ( 120 , 0.1987 )
            ( 150 , 0.1868 )
            ( 180 , 0.1668 )
            ( 210 , 0.1634 )
			};	
		\addplot[color=7,
		mark=*,
		line width=2pt,
		smooth] coordinates {
			( 60 , 0.1693)
			( 90 , 0.1793)
            ( 120 , 0.1997)
            ( 150 , 0.1900 )
            ( 180 , 0.1897 )
            ( 210 , 0.1801 )
		};
			\legend{CAL, PHO, SIN, NY}
		\end{axis}
		\end{tikzpicture}
	}
 	\vspace{-0.1in}
	\caption{Impacts of hyper-parameters on MCMG regarding HR@10 and NDCG@10.
	}\label{fig:parameter_sensitivity}
 	\vspace{-0.1in}
\end{figure*}

%% file: section/conclusion.tex
\section{CONCLUSIONS}
This paper, \textit{for the first time}, proposes a multi-channel next POI recommendation framework with multi-granularity check-ins signals (MCMG). Guided by the extensive data analysis, MCMG is equipped with three modules. In particular, the global user behavior encoder aims to capture the augmented sequential regularity from the global check-in signals for more expressive POI representations; the local multi-channel encoder composed of location, region and category encoders seeks to mine the local check-ins signals on the basis of the global user behavior encoder; and the region-aware weighting strategy dynamically aggregates the impact of the three local channels to ease the potential contradictory prediction issue, thus generating more accurate next POI recommendation. 
Experimental results on four real-world datasets demonstrate the superiority of our MCMG against state-of-the-art next POI recommendation approaches.